\documentclass[lettersize,journal]{IEEEtran}
\usepackage{amsmath,amsfonts}
\usepackage{algorithm, algpseudocode}
\usepackage{multicol}

\usepackage{array}
\usepackage[caption=false,font=normalsize,labelfont=sf,textfont=sf]{subfig}

\newcolumntype{C}{>{\centering\arraybackslash}p{0.5cm}}
\usepackage{textcomp}
\usepackage{stfloats}
\usepackage{multirow}
\usepackage{url}
\usepackage{nicematrix}
\usepackage{mathrsfs}
\usepackage{verbatim}
\usepackage{graphicx}
\usepackage{xcolor}
\usepackage{listings}
\definecolor{codegreen}{rgb}{0,0.6,0}
\definecolor{codegray}{rgb}{0.5,0.5,0.5}
\definecolor{codepurple}{rgb}{0.58,0,0.82}
\definecolor{backcolour}{rgb}{0.95,0.95,0.92}

\lstdefinestyle{mystyle}{
    backgroundcolor=\color{backcolour},   
    commentstyle=\color{codegreen},
    keywordstyle=\color{magenta},
    numberstyle=\tiny\color{codegray},
    stringstyle=\color{codepurple},
    basicstyle=\ttfamily\footnotesize,
    breakatwhitespace=false,         
    breaklines=true,                 
    captionpos=b,                    
    keepspaces=true,                 
    numbers=left,                    
    numbersep=5pt,                  
    showspaces=false,                
    showstringspaces=false,
    showtabs=false,                  
    tabsize=2
}
\begin{document}

\title{Detecting Code Vulnerabilities with Heterogeneous GNN Training}


\author{Yu Luo, Weifeng Xu, Dianxiang Xu

\thanks{...}
\thanks{Manuscript received April XX, 2023; revised XX XX, 202X.}}

\markboth{Journal of \LaTeX\ Class Files,~Vol.~14, No.~8, August~2021}%
{Shell \MakeLowercase{\textit{et al.}}: A Sample Article Using IEEEtran.cls for IEEE Journals}


\maketitle

\begin{abstract}
Detecting vulnerabilities in source code is a critical task for software security assurance. Graph Neural Network (GNN) machine learning can be a promising approach by modeling source code as graphs. Early approaches treated code elements uniformly, limiting their capacity to model diverse relationships that contribute to various vulnerabilities. Recent research addresses this limitation by considering the heterogeneity of node types and using Gated Graph Neural Networks (GGNN) to aggregate node information through different edge types. However, these edges primarily function as conduits for passing node information and may not capture detailed characteristics of distinct edge types. This paper presents Inter-Procedural Abstract Graphs (IPAGs) as an efficient, language-agnostic representation of source code, complemented by heterogeneous GNN training for vulnerability prediction. IPAGs capture the structural and contextual properties of code elements and their relationships. We also propose a Heterogeneous Attention GNN (HAGNN) model that incorporates multiple subgraphs capturing different features of source code. These subgraphs are learned separately and combined using a global attention mechanism, followed by a fully connected neural network for final classification. The proposed approach has achieved up to 96.6\% accuracy on a large C dataset of 108 vulnerability types and 97.8\% on a large Java dataset of 114 vulnerability types, outperforming state-of-the-art methods. Its applications to various real-world software projects have also demonstrated low false positive rates.
\end{abstract}


\section{Introduction}

\IEEEPARstart{W}{ith} the widespread of computer technology, cybersecurity has become an increasing concern. 
As of January 2023, over 140,000 CVEs \cite{CVE} (Common Vulnerabilities and Exposures) were reported to the National Vulnerability Database \cite{NVD}. The number of CVEs has been steadily increasing over the years, with an average growth rate of around 15-20\% per year. Detecting vulnerabilities in code is a complex task that involves analyzing the codebase's structure, syntax, and semantics to identify potential weaknesses that attackers could exploit. Recent studies suggest that Graph Neural Network (GNN) machine learning is a promising approach \cite{cheng2021deepwukong} \cite{hin2022linevd} \cite{Li2018} \cite{liu2021combining}  \cite{Wang2020} \cite{zhou2019devign}. It models codebases as graphs with nodes representing code elements (such as functions, variables, and statements) and edges representing their relationships. By training the GNN on a large dataset of labeled code examples, the model can learn to identify patterns and features indicative of vulnerabilities that may be missed by other analysis techniques.

Early approaches have primarily focused on homogeneous GNN training, where all code element nodes and edges are treated uniformly. While homogeneous GNNs have shown potential in predicting vulnerabilities, they have limited capacity to model diverse relationships between code elements that contribute to various vulnerabilities. In recent research, there has been a shift toward acknowledging the heterogeneity of node types, leveraging Gated Graph Neural Networks (GGNN) to aggregate node information through various edge types. Yet, while these edges primarily act as conduits for passing node information, they may not fully capture the detailed characteristics associated with distinct edge types.

In this paper, we present a novel approach to detecting code vulnerabilities using heterogeneous GNN training. Our contribution is twofold. First, we propose Inter-Procedural Abstract Graphs (IPAGs) as an efficient, language-agnostic representation of source code for GNN-based vulnerability prediction that aims to map source code features to vulnerability labels. IPAGs are well-suited for the classification task by capturing how the source code tokens of a given routine (i.e., method in an object-oriented language or function in a procedural language), its callees, and their structural and contextual properties contribute to the routine as a whole. Additionally, IPAGs combine sequence and aggregation structures of syntactic properties into higher-level abstractions,

Second, we propose a heterogeneous attention GNN model (HAGNN) for detecting vulnerabilities with IPAGs. It partitions each IPAG into six subgraphs, capturing various features of the source code, such as the abstract syntax properties and contextual dependencies of source code tokens, routine declarations, and routine calls. We incorporate the subgraphs into a heterogeneous GNN layer that contains multi-message passing units, with each subgraph being learned separately. Then, we identify vulnerability features from the subgraph matrices with the global attention mechanism and perform the classification with a fully connected neural network.

To evaluate the effectiveness of our approach, we have applied it to two large datasets in C and Java collected from various sources, including the National Vulnerability Database (NVD) \cite{NVD}, Software Assurance Reference Dataset (SARD) \cite{SARD}, and other publications \cite{luo2022Journal} \cite{Wang2020}. The C dataset consists of 74,978 vulnerable functions of 108 vulnerability types and 168,346 non-vulnerable functions, whereas the Java dataset comprises 37,350 vulnerable methods of 114 vulnerability types and 68,480 non-vulnerable methods. The results of our experiment indicate that our approach outperforms state-of-the-art techniques, achieving up to 96.6\% accuracy on the C dataset and 97.8\% on the Java dataset. The use of heterogeneous training resulted in lower error variance across multiple training runs. Furthermore, we applied the resulting models to real-world software projects, consisting of over 74,351 routines and 2,186 vulnerabilities, and achieved a high detection rate of 90\% with a low false positive rate.

The remainder of this paper is organized as follows. Section II introduces the IPAGs. Section III describes the heterogeneous training; Section IV presents experiment results; Section V applies the resultant models to real-world projects; Section VI reviews related work; Section VII concludes this paper. 

\section{Inter-Procedural Abstract Graphs}

The IPAG of a routine is a graph $\langle N, E\rangle$, where $N$ and $E$ are the sets of nodes and edges, respectively. The nodes are the same as those from the routine's and its callees' ASTs (abstract syntax graphs). They fall into three categories:

\begin{enumerate}
\item \textbf{Token Nodes ($N_t$).} They are the fundamental syntactic elements of a routine's source code, including keywords such as ``if", ``else", and ``for", operators like ``+", ``-", ``*", and ``/", identifiers that refer to variable and function names, literals, punctuation symbols, and parentheses.

\item \textbf{Property Nodes ($N_p$).} They represent abstract components (properties) of programming constructs in the source code, such as a name, an expression, or a statement, through a sequence or branch structure.

\item \textbf{Declaration Nodes ($N_d$).} They represent the declarations of the routine and its callees.  
\end{enumerate}

\begin{figure}[h]
   \centering
   \includegraphics[scale = 0.6]{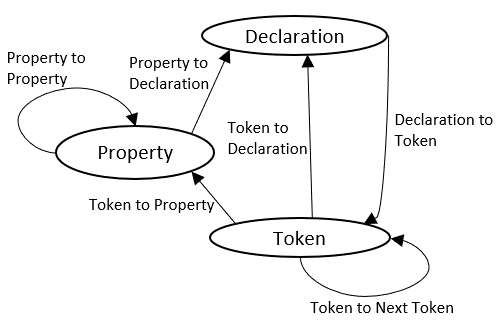}
   \caption{Nodes and Edges in IPAGs}
   \label{fig:HTGDATA}
\end{figure}

As shown in Fig. \ref{fig:HTGDATA}, there are six types of edges between the nodes. They are described below.

\begin{enumerate}

\item \textbf{Property to Declaration Edges ($E_{pd}$).} They represent the relationship between the declaration node and its associated property nodes, such as the modifiers, name, parameters, and return type of the routine.

\item \textbf{Property to Property Edges ($E_{pp}$).} They represent relationships between property nodes. A property node connected from another property node may form a sequence; A property node (e.g., function call expression) connecting from multiple property nodes (e.g., identifier expressions) may form an aggregation structure (i.e., the function call has multiple identifiers). 

\item \textbf{Token to Property Edges ($E_{tp}$).} Each edge connects a source code token to its syntactic property node.

\item \textbf{Token to Next Token Edges ($E_{tt}$).} Each edge connects a token node to its next token node. Collectively, the edges 
in $E_{tt}$ present the complete source code. 

\item \textbf{Token to Declaration Edges ($E_{td}$).} Each edge connects a token node to the declaration node of the routine to which the source code token belongs.   

\item \textbf{Callee Declaration to Caller Token Edge ($E_{dt}$).} Each edge $(dn, tn)$ in $E_{dt}$ represents a callee-caller relationship between the declaration node of a callee routine $dn$ and the token node of the caller $tn$. $E_{dt}$ includes an edge for each callee in the given routine. 
\end{enumerate}

\begin{figure*}[h]
    \centering
    \includegraphics[width=0.95\textwidth, height=8cm]{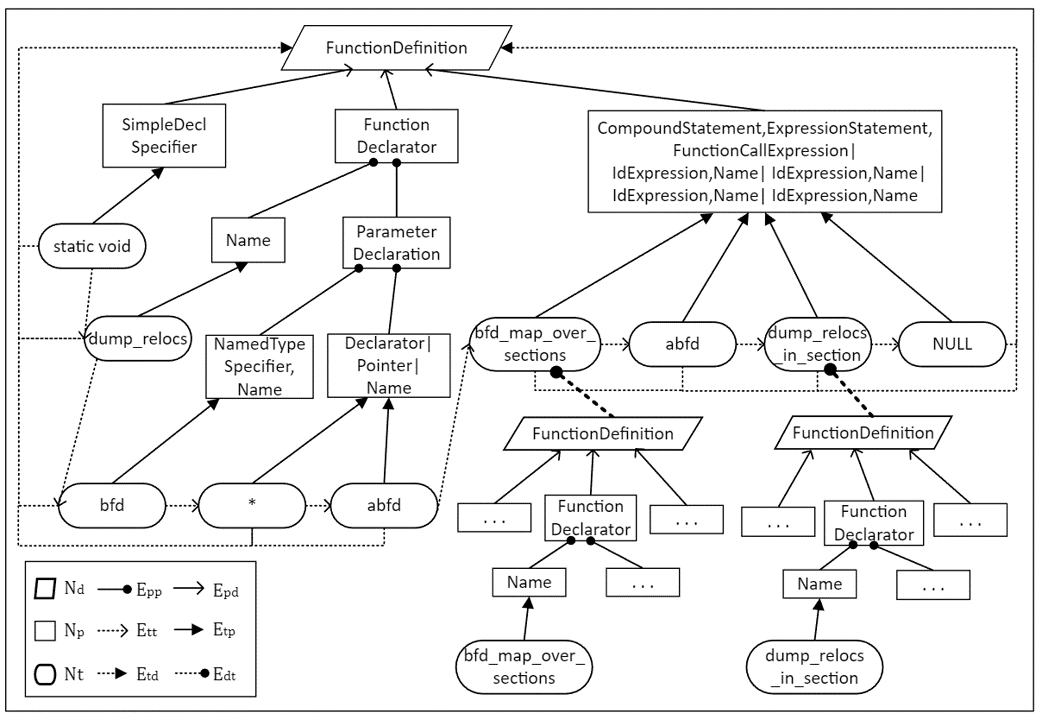}
    \caption{The IPAG of C Function dump\_relocs}
    \label{fig:IPAGEXAMPLE}
\end{figure*}

As such, we denote an IPAG as a 9-tuple $\langle N_t, N_p, N_d, E_{pd}, \newline E_{pp}, E_{tp}, E_{tt}, E_{td}, E_{dt}\rangle$. 
Fig. \ref{fig:IPAGEXAMPLE}
shows portion of the IPAG for function $dump\_relocs$, including the calls to  $bfd\_map\_over\_sections$ and  $dump\_relocs\_in\_section$. The C code of these functions is given in Listing 1. $dump\_relocs\_i\_section$ is subject to denial of service attacks 
due to the potential integer overflow in $reloc count$. Its calling functions $bfd\_map\_over\_sections$ and $dump\_relocs$ are also vulnerable. 

In Fig. \ref{fig:IPAGEXAMPLE}, the text in all token nodes, such as ``static void", ``dump\_relocs," and ``bdf", constitutes the source code. The token-to-property edges (``*", ``Declarator$|$Pointer$|$Name") and (``abfd", ``Declarator$|$Pointer$|$Name") indicate that ``*" is a pointer and  ``abfd" is a name. They form the variable declaration ``*abfd". The token-to-next-token edge (``*", ``abfd") ensures that the variable is ``*abfd" and not ``abfd*". 

A property node may be connected to an upper-level property that represents a higher level of syntactical abstraction. The three top-level property nodes ``SimpleDeclSpecifier", ``FunctionDeclarator", and ``CompoundStatement", represent the function modifier, declarator, and body, respectively. They form the essential components of a function through three property-to-declaration edges. In general, the path from a token node to the declaration node through one or more property nodes (e.g., ``abfd", ``Declarator$|$Pointer$|$Name", ``ParameterDeclaration", ``FunctionDeclarator", ``FunctionDefinition") is a sequence of increasing abstractions. It illustrates how the source code token (e.g., ``abfd") contributes to the function declaration as a whole. Additionally, the ``FunctionCallExpression" comprises four token nodes: ``bfd\_map\_over\_sections", ``abfd", ``dump\_relocs\_in\_section", and ``NULL". Of these, ``bfd\_map\_over\_sections" and ``dump\_relocs\_in\_section" are two callee functions and have their declaration nodes connected to related token nodes in the caller function (dump\_relocs) through the edges (``FunctionDefinition", ``bfd\_map\_over\_sections") and (``FunctionDefinition", ``bfd\_map\_over\_sections"), respectively.

\begin{lstlisting}[language=C, caption= A Vulnerable Code Example (CVE-2017-17122)]
static void dump_relocs (bfd *abfd){
    bfd_map_over_sections (abfd, dump_relocs_in_section, NULL);}

void bfd_map_over_sections (bfd *abfd, void (*func) (bfd *, asection *, void *),void *data){
  bfd_boolean more_sections;
  asection *section;
  BFD_ASSERT (abfd != NULL);
  BFD_ASSERT (func != NULL);
  ...
  for (section = abfd->sections; section != NULL; section = section->next){
    dump_relocs_in_section (abfd, section, NULL);}
  ...}

static void dump_relocs_in_section (bfd *abfd, asection *section,void *dummy ATTRIBUTE_UNUSED){
    arelent **relpp;
    ...}
\end{lstlisting}

We construct a routine's IPAG in three steps: (1) creating the preliminary IPAGs of the routine and its callees from their ASTs. (2) compressing
the preliminary IPAGs by merging sequence and aggregation structures. (3) updating the routine's IPAG with additional edges from the declaration nodes of callee IPAGs to the corresponding calling tokens. IPAGs extend CAGs (Compact Abstract Graphs) in our prior work \cite{luo2022} by considering the ordering of source code tokens in step 1 and routine calls in step 3.   

\subsection{Building Preliminary IPAGs}

We create the preliminary IPAG of a routine from its AST as follows:

\begin{itemize}

\item Reverse the direction of each edge in the AST. It captures how the source code tokens constitute the properties of individual programming constructs. It aligns with the classification task that maps the tokens and programming constructs to a decision about the routine as a whole (i.e., represented by the routine's declaration node).  

\item Add an edge from each source code token to the root (i.e., the routine's declaration node). It captures the immediate relationship between each token and the routine's declaration.

\item Add an edge from each source code token to the next token if applicable. It captures the contextual relationships between the source code tokens, i.e., the routine is defined by the tokens in the given order. 

\end{itemize}

Let $\langle N, T, E, r \rangle$ be the AST, where $N$ is the set of non-terminal property nodes, $T$ is the set of terminal (token) nodes, $r$ $\in N$ is the root (declaration) node, ${E}$ is the set of directed edges between the nodes in $N \cup T$. The preliminary IPAG is defined by $\langle N_t, N_p, N_d, E_{pd}, E_{pp}, E_{tp}, E_{tt}, E_{td}, E_{dt}\rangle$, where: 

\begin{enumerate}

\item $N_t =T$: the token nodes are the AST's terminal nodes. 

\item $N_p$ = $N \setminus \{r\}$: the property nodes are the AST's non-terminal nodes except the root.

\item $N_d$=$\{r\}$: the declaration nodes include and only include the routine's AST root (the callees will be handled later).

\item $E_{pd}$ = $\{(x,r):(r,x)\in E\}$: the property-to-declaration edges reverse the edges from the AST root to its descendants.

\item $E_{pp}$ = $\{(x,y):(y,x)\in E$ for $x, y \in N \setminus \{r\}\}$: The property-to-property edges reverse the corresponding edges in the AST.  

\item $E_{tp}$ = $\{(y,x):(x,y)\in E$ for each $x \in N, y \in T\}$: The token-to-property edges reverse the corresponding edges in the AST.

\item $E_{tt}$ = $\{(x, y): x \in T, y \in T, y$ is the next token of $x\}$. The token-to-next-token edges include an edge from each token to its next.  

\item $E_{td}$ = $\{(x, r): x \in T\}$: The token-to-declaration edges include an edge from each token to the declaration node.

\item $E_{dt}$ = $\{\}$. The preliminary IPAG does not represent call relationships although the preliminary IPAGs of its callees are created separately. 

\end{enumerate}

\begin{figure*}[h]
    \centering
    \includegraphics[width=0.95\textwidth, height=8cm]{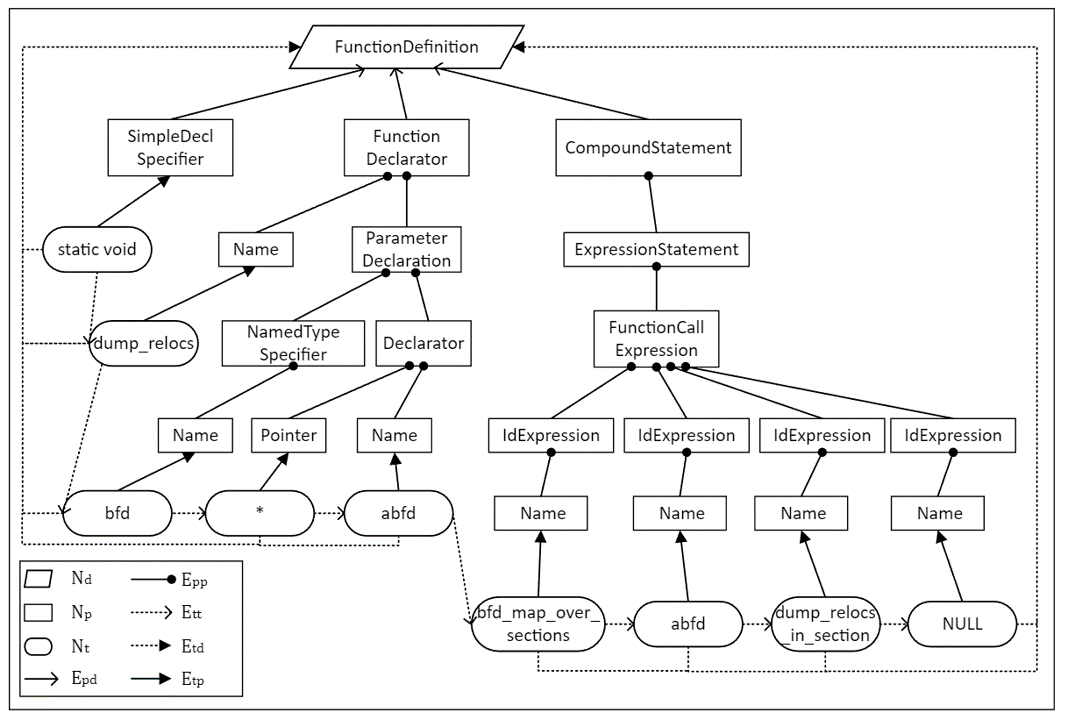}
    \caption{The Preliminary IPAG of C Function dump\_relocs}
    \label{fig:PIPAGEXAMPLE}
\end{figure*}

Fig. \ref{fig:PIPAGEXAMPLE} shows the preliminary IPAG of C function $dump\_relocs$. There are 9 token nodes of the source code, 20 property nodes, one declaration node, 3 $E_{pd}$ edges, 17 $E_{pp}$ edges, 9 $E_{tp}$ edges, 8 $E_{tt}$ edges, and 9 $E_{td}$ edges.

\subsection{Compressing Sequence and Aggregation Structures}

We reduce the preliminary IPAGs by merging property node sequences and compressible aggregation structures without losing information about the source code. It will make training more efficient due to the reduced graph sizes.

\subsubsection{Merging Property Node Sequences}
 
A property node sequence is defined as a list of property nodes $\langle n_1, n_2,...n_k \rangle$ ($k \geq 2$) that meets the following conditions: (1) $n_1$ is connected from a token node or one or more property nodes. (2) each $n_i \in N_p$ ($1<i\leq k$) has one entry and one exit edge, (3) $(n_i, n_{i+1})$ ($0<i<k$) $\in E_{pp}$, (4) the node connected from $n_k$ is either the declaration node or a property node with at least two entry edges. Such a sequence can be merged into one node without loss of information.


Consider the rightmost node sequence  $\langle$ ``Name'', ``IdExpression'' $\rangle$ in Fig. \ref{fig:PIPAGEXAMPLE}. The token node ``NULL'' connects to the sequence's first node ``Name''; whereas ``FunctionCallExpression'', the node connected from the sequence's last node, has four entry edges. The above property node sequence represents the abstractions that the source code token ``NULL'' is the name of an identifier in a function call expression. We merge these abstractions into one, label the new node by concatenating all node labels in the sequence, replace edge (``NULL'', ``Name'') with (``NULL'', new node), and edge (``IdExpression'', ``FunctionCallExpression'') with (new node, ``FunctionCallExpression''). Consider the top right node sequence $\langle$ ``FunctionCallExpression'', ``ExpressionStatement'', ``CompoundStatement'' $\rangle$ in Fig. \ref{fig:PIPAGEXAMPLE}. Four property nodes ``IdExpression'' connect to the sequence's first node ``FunctionCallExpression'', and the declaration node ``FunctionDefinition'' is connected from the sequence's last node.

\begin{figure*}[h]
    \centering
    \includegraphics[width=0.95\textwidth, height=8cm]{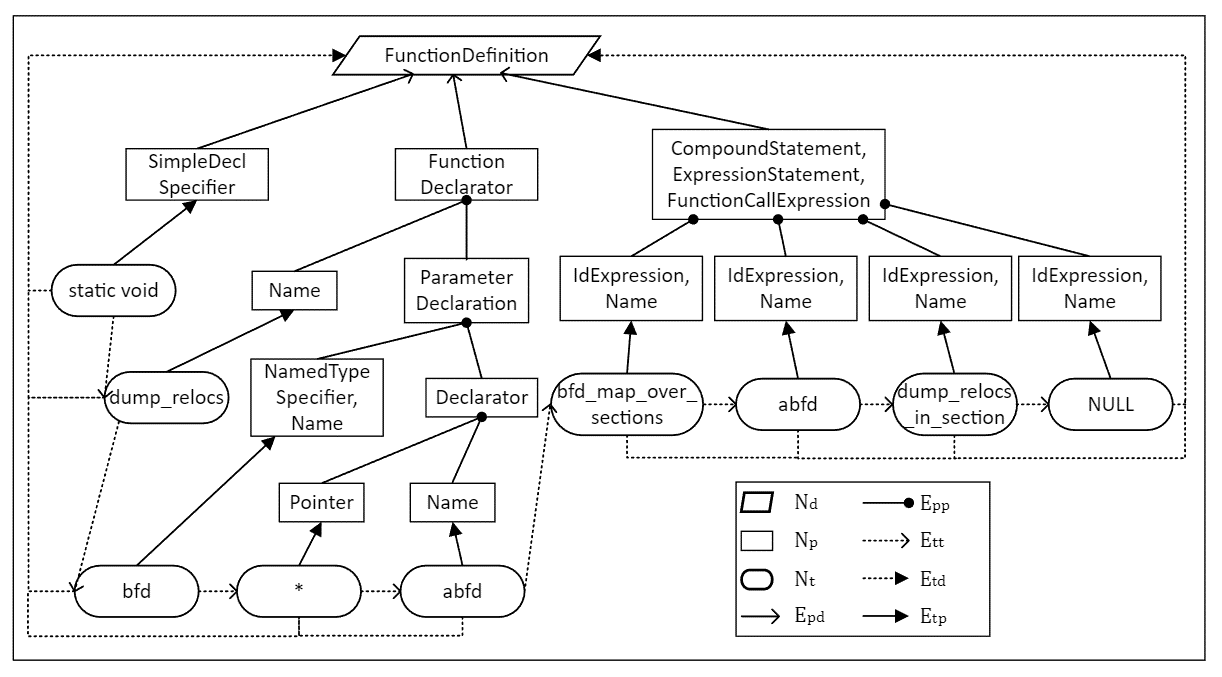}
    \caption{Sequence-Reduced IPAG of C Function dump\_relocs}
    \label{fig:CRIPAGEXAMPLE}
\end{figure*}

Given a preliminary IPAG $\langle N_t, N_p, N_d, E_{pd}, E_{pp}, E_{tp}, E_{tt}, \newline E_{td}, E_{dt}\rangle$, we compress the property node sequences in three steps: (1) Find all longest property node sequences. For each sequence denoted by $\{n_1, n_2,...n_k\}$, we remove node $n_1$ to $n_k$ from property node set $N_p$, and add new merged node $n$ to $N_p$. (2) Remove all edges between nodes in $\{ n_1, n_2,...n_k \}$ and the edge $(n_k, \omega)$ from $E_{pp}$, and add the edge $(n, \omega)$ to $E_{pp}$, where $\omega$ is the node connected by the last sequence's node $n_k$ . (3) If the first sequence's node $n_1$ is connected by a token node $\alpha$, edge $(\alpha, n_1)$ in token edge set $E_{tp}$ is replaced by edge $(\alpha, n)$. If $n_1$ is connected by a set of property nodes, for each node $\alpha$ in this set, replace $(\alpha, n_1)$ by $(\alpha, n)$ in $E_{pp}$. After compression, the six node sequences in Fig. \ref{fig:PIPAGEXAMPLE} are merged into a ``NamedTypeSecifier, Name", a ``CompoundStatement, ExpressionStatement, FunctionCallExpression", and four ``IdExpression, Name" in Fig. \ref{fig:CRIPAGEXAMPLE}.

\subsubsection{Merging Aggregation Structures}

An aggregation structure is a group of property nodes $\langle \mu, n_1, n_2,...n_k \rangle$ ($k \geq 2$), where $\mu$ is the ``parent'', and its children are $n_1, n_2,...n_k$. It satisfies five conditions: (1) $\mu \in N_p$ is a property node or sequence merged node with two or more entry edges. (2) Each $n_i \in N_p'$ ($0<i\leq k$) is a property or sequence merged node connecting to $\mu$. (3) $(n_i, \mu)$ ($0<i\leq k$) is an edge in $E_{pp}$. (4) $\{\theta_1, \theta_2, ..., \theta_k \}$ ($k \geq 2$) a list of property or token node connecting to $n_i$, for each $\theta_i$ ($0<i\leq k$), $(\theta_1, n_i) \in E_{pp} \cup E_{tp}$. (5) $\xi$, the node connected from node $\mu$ is either the declaration node or a property node with one or more entry edges. 

Not all aggregation structures are compressible, depending on structural and semantic constraints. 

\begin{figure}[ht]
\begin{tabular}{ccccc}
\includegraphics[scale=0.4]{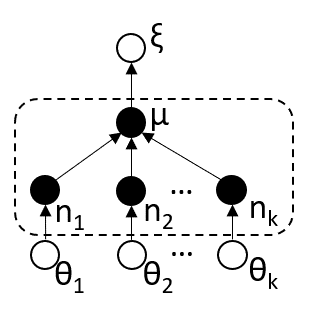}
&
\includegraphics[scale=0.4]{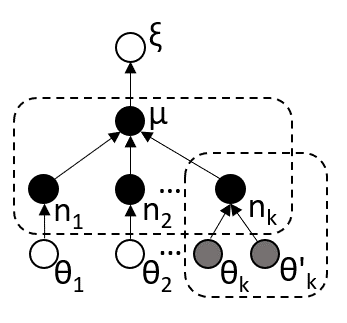} \\
(a)   & (b) \\
\end{tabular}
\caption{Sample Aggregation Structures}
\label{fig:aggregation}
\end{figure}

\textbf{Structural constraint.} Fig. \ref{fig:aggregation} shows two patterns of aggregation structures, where (a) is compressible, but (b) is not compressible. In (a), each child node $n_i$ in the structure has exactly one entry edge from $\theta_i$ (either a token node or property node). An example in Fig. \ref{fig:PIPAGEXAMPLE} is the parent node ``FunctionCallExpression'' together with its four child nodes of ``IdExpression". Fig. \ref{fig:aggregation} (b) is not compressible because the last child node $n_k$ is connected from two or more property nodes. Merging such an aggregation structure into one node would lose information in that $n_k$ has a different structure than $n_i$ ($i<k$). Consider the aggregation in the middle of Fig. \ref{fig:PIPAGEXAMPLE}: the parent node ``ParameterDeclaration" is connected from two child nodes, ``NamedTypeSpecifier" and ``Declarator". The child node ``Declarator" is connected from two property nodes (``Pointer" and ``Name"). This structure should not be reduced.

\textbf{Semantics constraint.} The parent node name of an aggregation structure indicates what the aggregation means. Only the aggregation with a whole-part relation between the parent and its children can be merged to present a meaningful higher-level abstraction -- the children constitute the programming construct indicated by the parent node. Consider the rightmost aggregation in Fig. \ref{fig:PIPAGEXAMPLE}. Four property nodes ``IdExpression" construct the complete function call expression that corresponds to the source code ``bfd\_map\_over\_sections(abfd, dump\_relocs\_in\_section, NULL);" in line 2 of listing 1.

We have identified the property names that represent whole-part relations in C and Java, respectively. For C, 33 of the 57 property names may appear in the parent node of a compressible aggregation. For Java, 45 of 63 property names may be the parent node. The property names fall into seven categories: expression, statement, declaration, argument, parameter, type, and specifier/modifier, as shown in Table \ref{table:AggregationCategory}, where '*' indicates that the aggregation structure is compressible. 

\begin{table*}
\caption{Categories of Aggregation Structures}
\begin{center}
\begin{tabular}{|l|l|l|}
\hline
\textbf{Category}&\textbf{Names of the Parent Property Node in C} & \textbf{Names of the Parent Property Node in Java}\\
\hline
Expression* & BinaryExpression, FunctionCallExpression, Array- & MethodCallExpr, FieldAccessExpr, BinaryExpr, Object-\\ 
& SubscriptExpression, CastExpression, Conditional- & CreationExpr, AssignExpr, ArrayCreationExpr, CastExpr,\\ 
& Expression, CompoundStatementExpression, Id- & ArrayAccessExpr, IntegerLiteralExpr, UnaryExpr, Instance-\\ 
& Expression, LiteralExpression, TypeIdExpression, & OfExpr, SingleMemberAnnotationExpr, VariableDeclaration-\\ 
& UnaryExpression & Expr, ConditionalExpr\\

\hline

Statement & CompoundStatement, IfStatement, ForStatement, & statements, IfStmt, CatchClause, TryStmt, SwitchEntry,\\
& DoStatement, SwitchStatement, WhileStatement, & ForStmt, WhileStmt, catchClauses, entries, SwitchStmt,\\
& LabelStatement, BreakStatement, Expression- & ExpressionStmt, ForEachStmt, LabeledStmt, DoStmt,\\
& Statement, ReturnStatement*, DeclarationStatement, &  AssertStmt, ThrowStmt, ReturnStmt*, thrownExceptions,\\ 
& ContinueStatement, TryBlockStatement & SynchronizedStmt\\

\hline

Declaration* & SimpleDeclaration, Declarator, ArrayDeclarator, & VariableDeclarator, variables, values\\
& FunctionDeclarator, ParameterDeclaration & \\
\hline
Parameter/Initializer* &InitializerList & Parameter, parameters, typeParameters, TypeParameter\\
\hline
Type* & TypeId & ClassOrInterfaceType, ArrayType\\
\hline
Specifier/Modifier* & CompositeTypeSpecifier, SimpleDeclSpecifier & modifiers \\
\hline
Argument* & arguments & arguments, typeArguments\\
\hline
\end{tabular}

  \begin{minipage}{\textwidth}
    \small
    \vspace{0.1cm}
    \hspace{1.9em} *:  compressible
  \end{minipage}

\label{table:AggregationCategory}
\end{center}
\end{table*}

The structure of an expression consists of various components such as constants, variables, functions, and operators. When these components are combined to form an expression, the parent node is linked to all child nodes to represent the entire expression. However, it is possible to consolidate the expression into a single node while still retaining essential information. For example, the ``FunctionCallExpression'' node consists of four ``IdExpression, Name'' nodes. After merging, we use a node with the above elements to reflect the line of code ``bfd\_map\_over\_sections (abfd,
dump\_relocs\_in\_section, NULL);''. Therefore, all aggregations with labels in the expression structure can be compressed.

Unlike expressions, the child nodes in a statement aggregation (excluding the return statement) represent a parallel relationship between each line of source code tokens within the same block. Combining the nodes in a statement would result in the loss of information regarding their parallel connections. The return statement is a unique type of statement that is comparable to an expression, as all child nodes form the complete return statement. Statement aggregations, except for the return statement, cannot be combined.

The labels in the remaining categories in Table \ref{table:AggregationCategory} indicate that elements possessing similar attributes combine to form a unified entity. For instance, in C, the ``InitializerList" initializes data structures with multiple values, and in Java, the ``variabledeclarator" demonstrates the declaration of a variable. Hence, all aggregations categorized under ``Declaration", ``Parameter/Initializer", ``Type", ``Specifier/Modifier" and ``Argument" can be compressed. To avoid losing ordering and position information when merging multiple node labels, we incorporate relevant features into node embeddings as described in Section IV.B.

\begin{algorithm}[h] 
\small
\caption{Compression of Aggression Structures}
\label{alg:MergeAggregation}
\begin{algorithmic}[1]
\Require{a sequence reduced IPAG $\langle N_t, N_p, N_d, E_{pd}, E_{pp}, E_{tp},  E_{tt}, \newline E_{td}, E_{dt} \rangle$, a compressible aggregation structure extractor, $A()$} 
\Ensure{aggression reduced IPAG $\langle N_t, N_p, N_d, E_{pd}, E_{pp}, E_{tp}, E_{tt} \newline , E_{td}, E_{dt} \rangle$}
\Statex
\State {$L_a = A(\langle N_t, N_p, N_d, E_{pd}, E_{pp}, E_{tp}, E_{tt}, E_{td}, E_{dt} \rangle$)}
\For {$l \in L_a$}
    \State{$l = \{\mu, n_1, n_2,...n_k\}$}{\Comment{$\mu$ is the parent node, $n_1$ to $n_k$ are $\newline {\hspace{11.3em}}$ child nodes in the aggregation}}
    \State {$N_p' \leftarrow N_p \cup \{m\} \setminus \{\mu, n_1, n_2, ..., n_k\}$}{\Comment{$m$ is the new $\newline {\hspace{20.1em}}$ merged node}}
    
    \State {$E_{pp}' \leftarrow E_{pp} \cup \{(m,x)\}  \setminus \{(\mu,x)\} \setminus \{(n, \mu) : n \in \{n_1, n_2, \newline {\hspace{1.7em}} ..., n_k\} \}$, where $x$ is the node connected by $\mu$, $(\mu,x) \in E_{pp}$}

    \If{$\{y : (y, n) \in E_{tp}$ for each $n \in \{n_1, n_2, ..., n_k\}\} \in N_t$}
        \State {$E_{tp}' \leftarrow E_{tp} \cup \{(y,m) : \forall(y,n) \in E_{tp}$ for each $n \in \{n_1, \newline {\hspace{3.1em}}  n_2, ..., n_k\} \setminus \{(y,n) : \forall(y,n) \in E_{tp}$ for each $n \in \{n_1,\newline {\hspace{3.1em}} n_2, ..., n_k\} \}$}
    \Else
        \State {$E_{pp}' \leftarrow E_{pp}' \cup \{(y,m) : \forall(y,n) \in E_{pp}'$ for each $n \in \{n_1, \newline {\hspace{3.1em}} n_2, ..., n_k\} \setminus \{(y,n) : \forall(y,n) \in E_{pp}'$ for each $n \in \{n_1, \newline {\hspace{3.1em}} n_2, ..., n_k\}\}$}
    \EndIf
\EndFor
\State $N_p = N_p'$
\State $E_{pp} = E_{pp}'$
\State $E_{tp} = E_{tp}'$
\State {Return aggression reduced IPAG $\langle N_t, N_p, N_d, E_{pd}, E_{pp}, E_{tp}, \newline E_{tt}, E_{td}, E_{dt} \rangle$}
\end{algorithmic}
\end{algorithm}

\begin{figure*}[h]
    \centering
    \includegraphics[width=0.95\textwidth, height=8cm]{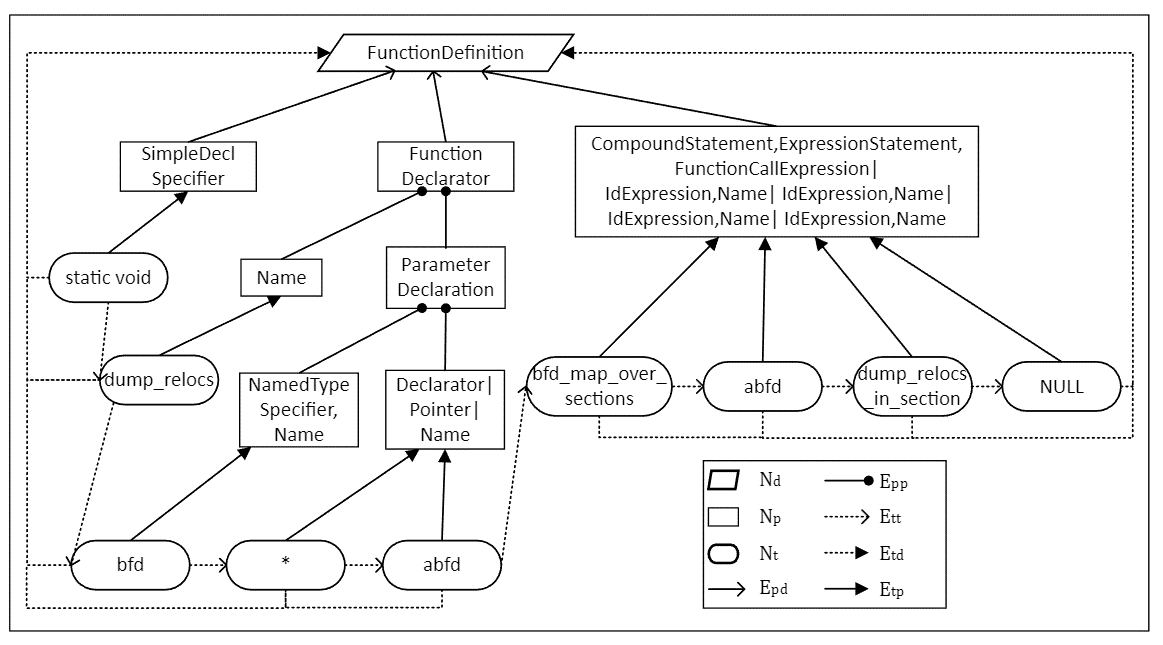}
    \caption{Aggregation-Reduced IPAG of C Function dump\_relocs}
    \label{fig:CAGEXAMPLE}
\end{figure*}

Algorithm \ref{alg:MergeAggregation}  describes the process, where function $A()$  algorithm obtains all compressible aggregations. For each aggregation structure $l$= $\{\mu, n_1, n_2,...n_k\}$, where $\mu$ is parent node and $n_1$ to $n_k$ are child nodes in the aggregation, the algorithm removes all nodes in $\langle \mu, n_1, n_2,...n_k \rangle$ from property node set $N_p$, and adds new merged node $m$ to $N_p$ (\textbf{line 4}). Then it removes all edges from child nodes in $\langle n_1, n_2,...n_k \rangle$ to parent node $\mu$ and the edge between the parent node $\mu$ and the property node $x$ from property edge set $E_{pp}$, and adds the edge $(m, x)$ to $E_{pp}$, where $m$ is the new merged node (\textbf{line 5}). In \textbf{lines 6-10}, if each child node $n$ is connected by a token node $y$, edge $(y, n)$ in the token edge set $E_{tp}$ is replaced by edge $(y, m)$. If each child node $n$ is connected by a property node $y$, edge $(y, n)$ in the property edge set $E_{pp}'$ is replaced by edge $(y, m)$. The aggression-reduced IPAG of function dump\_relocs in listing 1 is shown in Fig. \ref{fig:CAGEXAMPLE}. Among the four aggregation structures, two are compressed (i.e., ``FunctionCallExpression'' and ``Declarator''), whereas ``FunctionDeclarator'' and ``ParameterDeclaration'' are not compressible.

\subsection{Processing Call Relationships}

To establish call relationships, we connect the declaration nodes of all traceable callees' IPAGs to the corresponding token nodes (callee names) of the caller's IPAG (a callee is traceable if its source code is available in the given dataset). Algorithm \ref{alg:BuildCall} describes the process, where $G$ is the aggression-reduced IPAG set for the entire dataset, $C$ is a function that returns the list of traceable callees in the given routine, and $D$ is the call relation depth of a routine.

\begin{algorithm}[h]
\small
\caption{Building Call Relations}
\label{alg:BuildCall}
\begin{algorithmic}[1]
\Require{aggregation-reduced IPAG set $G$ for the entire dataset}
\Ensure{the complete IPAG set $G'$ for the entire dataset}
\Statex
\State {$\{G_0, G_1, ... G_n\} \leftarrow G_i = \{g \in G \mid D(g) = i\}, \quad 0 \leq i \leq n$, where n is the maximum value of the deepest depth}
\State {$G' \leftarrow G_0$}
\For {$i \in [1:n]$}
    \For {each IPAG $\langle N_t, N_p, N_d, E_{pd}, E_{pp}, E_{tp}, E_{tt}, E_{td}, E_{dt} \rangle \newline {\hspace{2.9em}} \in G_i$}
        \State {$\{c_1, c_2, ..., c_k\} \leftarrow \{c \in \{$``FunctionCallExpression", $\newline {\hspace{3.1em}}$``MethodCallExpr"$\} : c \in N_p\}$, $k \geq 0$}
        \If{$k \neq 0$}
            \State{$\{t_1, t_2, ..., t_l\} \leftarrow \{t : (t,c) \in E_{tp}\}$ for each $c \in \newline {\hspace{4.5em}} \{c_1, c_2, ..., c_k\}$}
            \State{$\{\tau_1, \tau_2, ..., \tau_j\} \leftarrow  C(\{t_1, t_2, ..., t_l\}),0 \leq j \leq l$}
            \If{$ j \neq 0$}
                \State{$\{d_1, d_2, ..., d_j\} \leftarrow \{d : d \in \{G_1', G_2', ..., G_j'\}\}\newline {\hspace{1.1em}}$ {\Comment{$ G_j'$ is the complete IPAG of callee $\tau_j$}}}
                \State{$N_t' \leftarrow N_t \cup \{\mathcal{N}_t : \mathcal{N}_t \in \{G_1', G_2', ..., G_j'\}\}$}
                \State{$N_p' \leftarrow N_p \cup \{\mathcal{N}_p : \mathcal{N}_p \in \{G_1', G_2', ..., G_j'\}\}$}
                \State{$N_d' \leftarrow \{d_1, d_2, ..., d_j\}$}
                \State{$E_{pd}' \leftarrow E_{pd} \cup \{\mathcal{E}_{pd} : \mathcal{E}_{pd} \in \{G_1', G_2', ..., G_j'\}\}$}
                \State{$E_{pp}' \leftarrow E_{pp} \cup \{\mathcal{E}_{pp} : \mathcal{E}_{pp} \in \{G_1', G_2', ..., G_j'\}\}$}
                \State{$E_{tp}' \leftarrow E_{tp} \cup \{\mathcal{E}_{tp} : \mathcal{E}_{tp} \in \{G_1', G_2', ..., G_j'\}\}$}
                \State{$E_{tt}' \leftarrow E_{tt} \cup \{\mathcal{E}_{tt} : \mathcal{E}_{tt} \in \{G_1', G_2', ..., G_j'\}\}$}
                \State{$E_{td}' \leftarrow E_{td} \cup \{\mathcal{E}_{td} : \mathcal{E}_{td} \in \{G_1', G_2', ..., G_j'\}\}$}
                \State{$E_{dt}' \leftarrow \{(d_i, \tau_i) : i \in [1:j]\}$}
            \EndIf
        \EndIf
        \State{$G' \leftarrow G' \cup \langle N_t', N_p', N_d', E_{pd}', E_{pp}', E_{tp}', E_{tt}', E_{td}', E_{dt}' \rangle$}
    \EndFor
\EndFor

\State {Return $G'$}
\end{algorithmic}
\end{algorithm}

\textbf{Step 1 (line 1-4).} We begin by determining the depth of the deepest call relation in each routine's aggression-reduced IPAG in the set $G$. Next, we partition the set $G$ into multiple sub-lists based on the deepest depth observed in each graph. The resulting sub-lists $G_0, G_1, \ldots, G_n$ collectively represent the division of the list $G$ according to the depth of each graph. For graphs in the sub-list $G_0$ that don't have any call relations, the aggression-reduced IPAG is equivalent to the complete IPAG. Moving on, we utilize a for loop to initiate the processing of call relations for graphs with a call relation depth of 1 and continue until graphs with a call relation depth of $n$. Since a routine can be called multiple times at different levels, and a graph with a higher call relation depth must call a graph with a lower call relation depth that has already been processed, effectively preventing duplication of work.

\textbf{Step 2 (line 5).} For each aggression-reduced IPAG,  locating all call-related property nodes $\{c_1, c_2, ..., c_k\} \in N_p$ $(k \geq 0)$ through the node name (``FunctionCallExpression" in C and ``MethodCallExpr" in Java). When $k = 0$, the routine has no call relation; the aggression-reduced IPAG is its complete IPAG.

\textbf{Step 3 (line 6--8).} If $k \neq 0$, extract all token nodes $\{t_1, t_2, ..., t_l\} \in N_t (l \geq k)$ connect to each call-related property node. $\{t_1, t_2, ..., t_l\} = \{t : (t,c) \in E_{tp}$ for each $c \in \{c_1, c_2, ..., c_k\} \}$. For each token node $t \in \{t_1, t_2, ..., t_l\}$, it could be three possibles: (a) a non-routine-name token (variable name, number, string, etc.), (b) a name of the routine that doesn't appear in the program (a routine from libraries), (c) a name of the routine that appears in the program (a routine from the same file or other files). We utilize the trackable callee checker to remove (a) and (b) from the above token node set resulting in a new set $\{\tau_1, \tau_2, ..., \tau_j\} (0 \leq j \leq l)$. When $j = 0$, the routine has no trackable call relation in the program; the aggression-reduced IPAG is its complete IPAG.

\textbf{Step 4 (line 9--21).} If $j \neq 0$, $\{d_1, d_2, ..., d_j\}$ are the declaration nodes in the complete IPAGs $\{G_1', G_2', ..., G_j'\}$ of those $j$ routines, whose name in $\{\tau_1, \tau_2, ..., \tau_j\}$. $N_t', N_p', N_d', E_{pd}', E_{pp}', E_{tp}', E_{tt}',  E_{td}'$ are updated by combining the relevant node and edge sets in caller's IPAG and all callees' IPAG, $\{G_1', G_2', ..., G_j'\}$. The new declaration node set $E_{dt}' = \{(d_i, \tau_i) : i \in [1:j]\}$ is the connection from each declaration node in callee's IPAG to its related token node in caller's IPAG.

To facilitate subsequent embedding and training, we use $\langle N_t, N_p, N_d, E_{pd}, E_{pp}, E_{tp}, E_{tt}, E_{td}, E_{dt}\rangle$ to represent the complete IPAG. The IPAG of the C function $dump\_relocs$ is shown in Fig. \ref{fig:IPAGEXAMPLE}. It has one ``FunctionCallExpression" connected by four token nodes (``bdf\_map\_over\_sections", ``abfd", ``dump\_relocs\_in\_section" and ``NULL"). Among them, ``bdf\_map\_over\_sections" and ``dump\_relocs\_in\_section" are two traceable routines, thus, their IPAGs respectively connect their related token nodes.

\section{Heterogeneous GNN Training for Vulnerability Detection}

\begin{figure*}[h]
    \centering
    \includegraphics[width=0.95\textwidth, height=8cm] {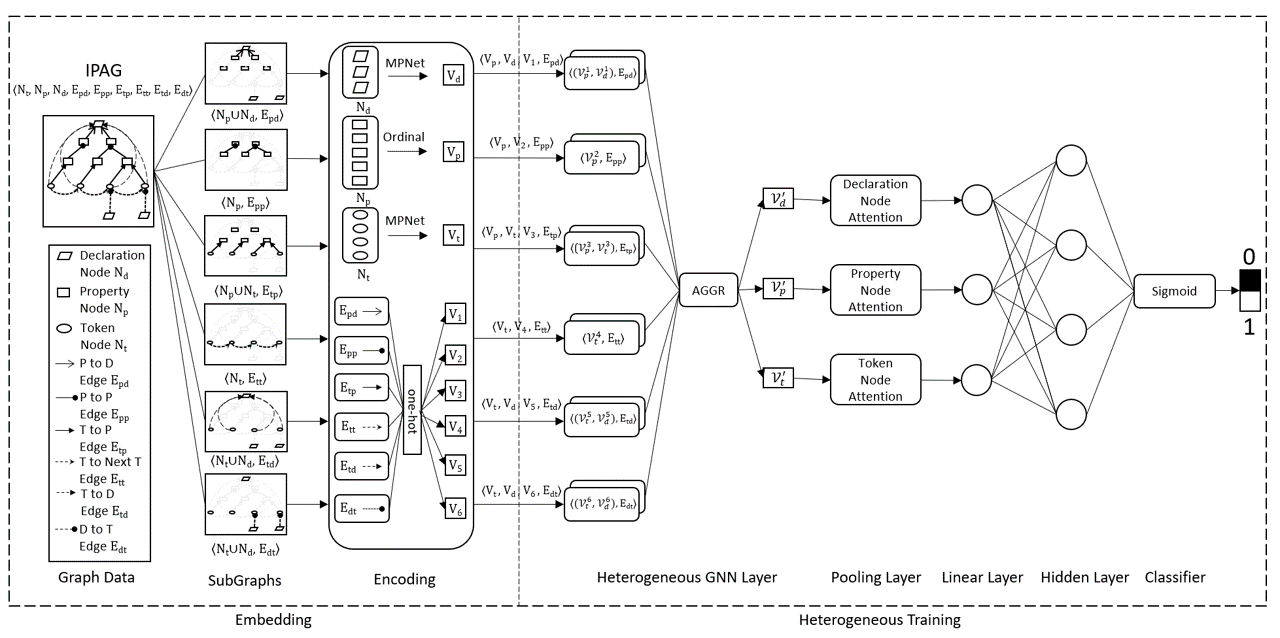}\\
    \caption{The Architecture of Heterogeneous Attention GNN Model (HAGNN) 
    for Vulnerability Detection}
    \label{fig:HGNN}
\end{figure*}

Fig. \ref{fig:HGNN} shows the architecture of heterogeneous attention GNN model (HAGNN) for vulnerability detection. It consists of an embedding phase and a heterogeneous training phase. The embedding phase involves two steps. First, the given IPAG $\langle N_t, N_p, N_d, E_{pd}, E_{pp}, E_{tp}, E_{tt}, E_{td}, E_{dt}\rangle$ is sliced into the following six subgraphs: 

\begin{enumerate}
 
\item $\langle N_p \cup N_d, E_{pd}\rangle$: the declaration subgraph of all function calls.

\item $\langle N_p, E_{pp}\rangle$: the subgraph of all property nodes of the source code tokens. 

\item $\langle N_t \cup N_p, E_{tp}\rangle$: the subgraph of all source code tokens and their immediate property nodes.  

\item $\langle N_t, E_{tt}\rangle$: the subgraph of all source code tokens and their contextual dependencies.  

\item $\langle N_t \cup N_d, E_{td}\rangle$: the subgraph of all source code tokens and their routine declaration nodes, 

\item $\langle N_d \cup N_t, E_{dt}\rangle$: the subgraph on function calls. 

\end{enumerate}

Second, the node names and edges are encoded in numeral format. This involves converting the declaration node $N_d$ and token node $N_t$ to $V_d$ and $V_t$ through an MPNet \cite{song2020mpnet} respectively, converting property node $N_d$ into ordinal vectors $V_d$, and converting six types of edges into one-hot vectors $V_1, V_2, V_3, V_4, V_5, V_6$. The numerical representations of six subgraphs fit into the heterogeneous training.

In the heterogeneous training phase, the node vectors in each subgraph are concatenated by the related edge vector and sent to a message-passing unit with the adjacency matrix respectively. For instance, the input from the declaration subgraph $\langle N_p \cup N_d, E_{pd} \rangle$ to the first message-passing unit is  $\langle \mathcal{V}_p^1, \mathcal{V}_d^1, E_{pd}\rangle$, where $\mathcal{V}_d^1 = CONCAT(V_d, V_1)$, $\mathcal{V}_p^1 = CONCAT(V_p, V_1)$, and $E_{pd}$ is the adjacency matrix. Then, each message-passing unit aggregates information between nodes through the adjacency matrix and produces a new feature vector for each node. Subsequently, the aggregator (AGGR) combines the updated node feature vectors to generate three new node vectors: the updated declaration node vector $\mathcal{V}_d^{'}$, the updated property node vector $\mathcal{V}_p^{'}$, and the updated token node vector $\mathcal{V}_t^{'}$. They pass through the corresponding attention layers to locate high-impact features. Next, a fully connected neural network is applied to these feature vectors, and the resulting output is passed through a sigmoid function to make the classification decision.


\subsection{Embedding}

IPAGs have three types of nodes and six types of edges. To feed IPAGs into a heterogeneous GNN model, all nodes and edges should be embedded into numeric vectors.

\textbf{Node Embedding.} Node embedding is a numeric representation of the name of each node in the IPAG, which could be a token node, a property node, or a declaration node. According to the different emphases of the information expressed by each node type in the graph, we have two node embedding methods:

(a) The token nodes $N_t$ focus on expressing the semantic information of the source code in the routine and pass messages through the token to property edges $E_{tp}$, token to next token edges $E_{tt}$, token to declaration edges $E_{td}$ to property nodes $N_p$, token nodes $N_t$, and declaration nodes $N_d$ respectively. The declaration nodes $N_d$ mainly receive semantic information from token nodes $N_t$ and syntax information from property nodes $N_p$ and pass them to token nodes $N_p$ through declaration to token edges $E_{dt}$. For each node in $N_p$ and $N_t$, we utilize MPNet \cite{song2020mpnet}, a pre-trained model for text understanding, to convert the node label into a 768 fixed-length numeric vector. Mapping names to a high-dimensional space will make the initial features of names with similar semantics closer. For example, ``long'' and ``int'' are both variable types, so their embedding vectors are close in the space. MPNet combines two mainstream language models, masked language modeling (MLM) from BERT \cite{feng2020codebert} and permuted language modeling (PLM) from \cite{yang2019xlnet}. It has the best average performance on multiple text-based tasks. As there is no merge node present in both $N_t$ and $N_d$, MPnet is able to directly convert their names into embedding vectors $V_t$ and $V_d$, denoted as $V_t = MPNet(N_t)$ and $V_d = MPNet(N_d)$.

\begin{figure}[h]
    \centering
    \includegraphics[width=0.45\textwidth, height=4cm] {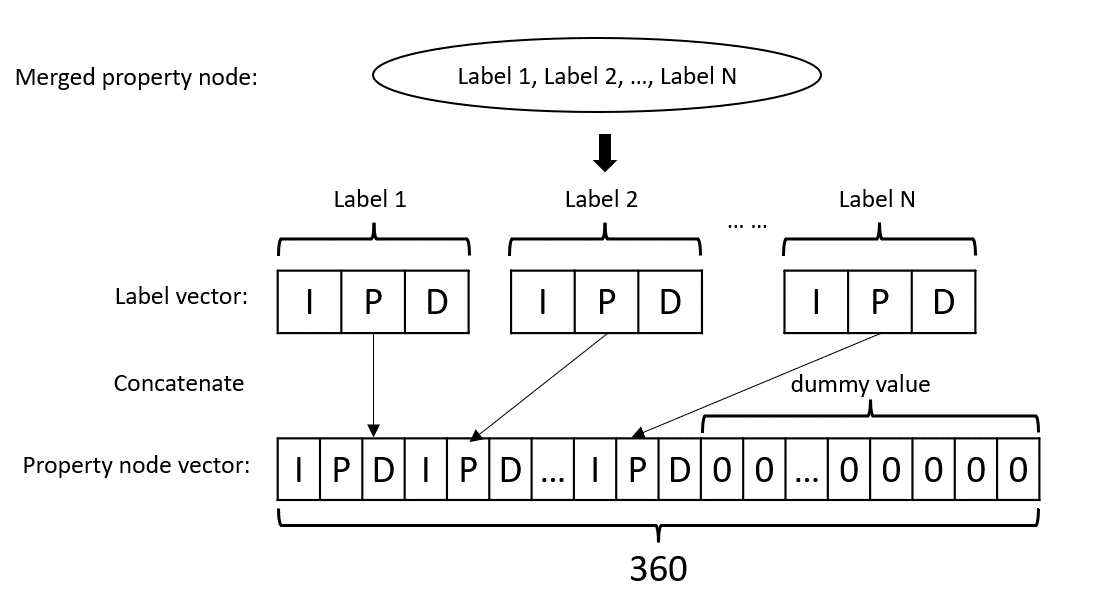}\\
    \caption{Property Node Embedding}
    \label{fig:NODEEMBEDDING}
\end{figure}

(b) The property nodes $N_p$ present the abstract syntax, which focuses more on property categories than semantic features. We use ordinal embedding to encode the property node name into a 360-fixed-length numeric vector. Fig. \ref{fig:NODEEMBEDDING} illustrates the property node embedding generation process. First, identify $N$ property node names in the dataset and assign each name a unique integer index in a vector with three elements: I is the name's index, P is the position of the name in an aggregation, and D is the depth of the name in a sequence. Given a node name, it could be three possibilities: (a) a name in a non-merged node. It is the only name in the node whose position and depth are both 1. So its embedded vector is $(I, 1, 1)$ concatenating with 357 dummy value 0. (b) a name in a sequence merged node. D is the name order in the sequence. And there is no aggregation in such a node, so $P = 1$. For example, given a sequence merged node with a name (a, b), its embedded vector is $(I_a, 1, 1, I_b, 1, 2)$ concatenating with 354 dummy value 0. (c) a name in an aggregation merged node. For names in the parent node, $P = 1$, and in a child node, P is the order number of the child node. For instance, given an aggregation merged node $(a\|b\|c,d)$, its embedded vector is $(I_a, 1, 1, I_b, 2, 1, I_c, 3, 1, I_d, 3, 2)$ concatenating with 348 dummy value 0. In our dataset, a merged node contains at most 104 names, meaning a node could have up to 312 features. We reserve 48 dummy values for special nodes that may appear in the application.

\textbf{Edge Embedding.} Edge embedding is used to represent the relationships between two nodes. We apply one-hot encoding to produce a 6-fixed-length numeric vector, where each bit represents a specific type of edge. The encoding scheme is as follows: $E_{pd}$ is (1, 0, 0, 0, 0, 0), $E_{pp}$ is (0, 1, 0, 0, 0, 0), $E_{tp}$ is (0, 0, 1, 0, 0, 0), $E_{tt}$ is (0, 0, 0, 1, 0, 0), $E_{td}$ is (0, 0, 0, 0, 1, 0), and $E_{dt}$ is (0, 0, 0, 0, 0, 1).

\subsection{Heterogeneous GNN Training}

To leverage the hidden knowledge from diverse node and edge types in IPAGs, we construct a heterogeneous attention graph neural network, where a single GNN layer contains multiple massage passing units for different types of edges. To embed IPAGs into a numerical format, we assign initial embedding vectors to declaration nodes and token nodes using a pre-trained MPnet, while property nodes are given ordinal vectors with position and depth features. The six types of edges are transformed into 6-fixed-length one-hot vectors. 

The proposed heterogeneous model is used to learn and aggregate information from different types of nodes and edges in IPAGs. It consists of two heterogeneous GNN layers, a pooling layer, a linear layer, a hidden layer, and a classifier.

\textbf{Heterogeneous GNN Layer.} The heterogeneous GNN layer is responsible for updating node features by gathering information from neighbors via the message-passing unit. Since the information propagated by different edges is different, we build six message-passing units to accept features of six subgraphs from the feature extraction stage, and nodes in each subgraph can communicate and exchange information with their neighbors, respectively. The six units have the same message-passing operation, but different weight matrices.

We propose a message-passing operation called SAGE$^+$, which incorporates edge depth into the embedding generation algorithm in SAGE \cite{hamilton2017inductive} to facilitate information exchange level by level in large source code-based graphs. Algorithm \ref{alg:MESSAGEPASSING} outlines this message-passing process. The input includes features for all nodes $v_m$, $\forall v_m \in V$, where $m$ is the node index, and $v_m$ is the input feature vector for node $m$. $K$ represents the number of layers, and $h_m^{(k)}$ denotes the embedding of the $m^{th}$ node in layer $k$. For the initial layer $k=0$, $h_m^{(0)}$ corresponds to the $m^{th}$ node's initial embedding, which is the input feature $v_m \in V$. At each layer $k$, the node embeddings are updated level by level by traversing the edge depths (from $D$ to $1$) using the following steps, where $d$ denotes the current edge depth, and $E_d$ is a set of edges at depth $d$. Firstly, for a node $j$, its aggregated message vector $a_j^{(k)}$ from all incoming edges at depth $d$ is computed using a MEAN aggregation function, where the neighboring nodes' vectors at the current layer $k$ are dot product with a learnable weight matrix $W_d^{(k)}$ at depth $d$ in layer $k$, $\forall (i,j) \in E_d$, where $n_i$ are the neighboring nodes of $j$. When $d=D$, the node vectors stay in layer $k-1$ without updates. For nodes connected with edges at depth $D$, their aggregated message vectors are obtained based on their neighboring nodes' vectors at the last layer $k-1$, which is represented as $h_i^{(k-1)}$. After computing the aggregated message vector $a_j^{(k)}$, the updated vector $h_j^{(k)}$ for node $j$ at layer $k$ is computed by concatenating its vector at the last layer $h_j^{(k-1)}$ with $a_j^{(k)}$, dot producting the result with a learnable matrix $W_c^{(k)}$, and passing through a ReLU activation function. This process is repeated $k$ times, and all node vectors at layer $K$ form the updated node feature matrix $\mathcal{V}$.

\begin{algorithm}[h]
\small
\caption{Message Passing Operation in SAGE$^+$}
\label{alg:MESSAGEPASSING}
\begin{algorithmic}[1]
\Require{Node feature matrix $V$, edge matrix $E$, number of layers $K$, depth of deepest edge $D$, weight matrices $W$}
\Ensure{Updated node feature matrix $\mathcal{V}$}
\Statex
\State{$h_m^{(0)} \leftarrow v_m, \forall v_m \in V$}{\Comment{$v_m$ is input feature vector for node $m$}}
\For{$k = 1...K$}
    \For {$d = D...1$}
        \If{$d = D$}
            \State{$a_j^{(k)} = \text{MEAN}\left(W_d^{(k)} h_i^{(k-1)} : \forall (i,j) \in E_d\right)$}
        \Else{}
            \State{$a_j^{(k)} = \text{MEAN}\left(W_d^{(k)} h_i^{(k)} : \forall (i,j) \in E_d\right)$}
        \EndIf
        \State{$h_j^{(k)} = \text{ReLU}\left(W_c^{(k)} \cdot (h_j^{(k-1)} \bigoplus a_j^{(k)})\right) : \forall j$}
    \EndFor
\EndFor

\State{Return $\mathcal{V} \leftarrow \{h_m^{(k)} : \forall v_m \in V$\}}
    
\end{algorithmic}
\end{algorithm}

The updated node features from different subgraphs will be aggregated into three matrices based on node types. For instance, given an IPAG $\langle N_t, N_p, N_d, E_{pd}, E_{pp}, E_{tp}, E_{tt}, E_{td}, E_{dt}\rangle$, the IPAG will be divided into six subgraphs, $\langle N_p \cup N_d, E_{pd}\rangle$, $\langle N_p, E_{pp}\rangle$, $\langle N_p \cup N_t, E_{tp}\rangle$,$\langle N_t, E_{tt}\rangle$,$\langle N_t \cup N_d, E_{td}\rangle$ and $\langle N_t \cup N_d, E_{dt}\rangle$. These subgraphs are then processed through a heterogeneous GNN layer, resulting in six fixed-length output sets, $\langle \mathcal{V}_d^1, \mathcal{V}_p^1\rangle$, $\langle \mathcal{V}_p^2\rangle$, $\langle \mathcal{V}_d^3, \mathcal{V}_t^3\rangle$, $\langle \mathcal{V}_p^4, \mathcal{V}_t^4\rangle$, $\langle \mathcal{V}_p^5\rangle$, $\langle \mathcal{V}_d^6, \mathcal{V}_t^6\rangle$. These outputs are then aggregated into three sets, which are denoted as $\mathcal{V}_d'$, $\mathcal{V}_p'$, and $\mathcal{V}_t'$.

\textbf{Pooling Layer.} We utilize the global soft attention (GSA) layer \cite{li2015gated} to assign a weight to each node, which aims to evaluate each node's contribution to the entire graph rather than just its local neighborhood. These weights will be learned during training and depend on the current state of the graph and the parameters of the heterogeneous GNN layer. Vulnerability detection is precisely to detect the impact of local code on the whole rather than the local. Therefore, the GSA can improve the performance of our heterogeneous model.

\textbf{Linear Layer, Hidden Layer, and Classifier.} A fully connected neural network is used to process three node feature sets obtained from the pooling layer. This network consists of a linear layer and a hidden layer that concatenates and converts the feature sets into a one-dimensional vector in the range $[0,1]$. The resulting vector is then used by a classifier to make a final decision, using a threshold of 0.5. If the output of the classifier is greater than 0.5, the routine is classified as vulnerable, otherwise, it is classified as non-vulnerable.

\section{EXPERIMENTS}


We have implemented the heterogeneous model with Pytorch v1.13.1 and PyG v2.2.0 and performed the experiments on a multi-core server with 4 Tesla V100S-PCIE GPUs. We use several common metrics to evaluate the performance of vulnerability detection, including accuracy, precision, recall, and F1 score. These measures are calculated based on the number of true positives ($TP$), true negatives ($TN$), false positives ($FP$), and false negatives ($FN$) predicted by the model. Accuracy is the percentage of all samples (both positive and negative) that the model correctly predicts, i.e., $(TP+TN)/(TP+TN+FP+FN)$. Precision is the proportion of positive samples that are correctly predicted out of all the samples predicted as positive, i.e., $TP/(TP+FP)$. Recall measures the proportion of actual positive samples that are correctly predicted, i.e., $TP/(TP+FN)$. The F1 score is the harmonic mean of precision and recall, providing a balanced measure of performance. It is defined as $2*(precision*recall)/(precision+recall)$. In summary, these metrics provide a comprehensive evaluation of the effectiveness of a vulnerability detection model by capturing different aspects of its performance.

We employed 5-fold cross-validation on the dataset of each experiment, allocating 80\% of the samples for classifier building (including both training and validation) and the remaining 20\% for testing. The initial epoch for each experiment was set at 50. If the accuracy and loss values were satisfactory after 50 epochs, we stopped the training. Otherwise, we continued with further training. Our reported evaluation metrics are based on the geometric mean across the 5 cross-validation folds.

\subsection{Datasets}

Table~\ref{table:Dataset} presents the datasets. The Java dataset originates from three sources: the national vulnerability database (NVD), the software assurance reference dataset (SARD), and Luo et al. \cite{luo2022Journal}. It has 37,350 positive samples with 114 vulnerability types and 68,480 negative samples. The C dataset, created from NVD, SARD, and Zhou et al. \cite{zhou2019devign}, contains 58,459 positive samples with 106 vulnerability types and 126,170 negative samples. The above vulnerabilities have covered 37 of the 40 general categories of software vulnerability. 

As mentioned in our previous work \cite{luo2022Journal}, the SARD dataset contains certain features that can introduce a significant bias in machine learning models, such as (1) logging statements before and after the vulnerable code that indicates the location of the vulnerability, and (2) comments that provide details about the vulnerability. To mitigate the impact of these features, we removed all comments and logging statements. The dataset and source code for reproduction are available at \cite{TIFS_HGNN_Dataset}.

\begin{table*}
\caption{The Datasets}
\begin{center}
\begin{tabular}{|l|c|c|c|c|c|c|c|c|}
\hline
 \textbf{Language} & \textbf{\#Vul.} & \textbf{\#Positive} & \textbf{\#Positive} & \textbf{\#Negative} & \textbf{\#Negative} & \textbf{Total} & \textbf{Total} & \textbf{Caller Sample} \\
 
\textbf{} & \textbf{Types} & \textbf{Samples}  & \textbf{Caller Samples} & \textbf{Samples} & \textbf{Caller Samples} & \textbf{Caller Samples} & \textbf{Samples} & \textbf{Ratio}\\

\hline
 Java &  114 & 37,350 & 12,198 & 68,480 & 27,375 & 39,573 & 105,830 & 37.39\% \\
\hline
 C & 108 & 74,978 & 18,016 & 168,346 & 65,486 & 83,502 & 243,324 & 34.32\%\\
\hline

\end{tabular}
\label{table:Dataset}
\end{center}
\end{table*}

\subsection{GNN Training with IPAGs}

Apart from our proposed SAGE$^+$, we evaluated five additional message-passing algorithms, implemented within the heterogeneous layer of the HAGNN framework: SAGE \cite{hamilton2017inductive}, GATs \cite{veličković2018graph}, UniMP \cite{shi2020masked}, RGGCN \cite{bresson2017residual}, and FiLM \cite{brockschmidt2020gnn}. These algorithms represent diverse approaches for handling graph representations. SAGE (GraphSAGE) uses the Laplacian matrix to perform aggregation operations and extract features from node neighborhoods. GAT utilizes a masked attention mechanism to aggregate neighbor node features, allowing the model to focus on specific nodes and edges that are most relevant to vulnerabilities. UniMP integrates both node features and label information to make predictions, enabling the model to leverage additional information about the nodes and their relationships for better predictions. RGGCN augments GCNs with gated margins and residuals, while FiLM uses a modulation mechanism to dynamically adjust its feature extraction.

\subsubsection{Overall Performance}

Table \ref{table:ResultC} presents the experimental results. For the C dataset, six models accomplished an accuracy rate exceeding 95\%, signifying the effectiveness of our framework. Among them, SAGE$^+$ showed the best performance on all performance metrics (accuracy, precision, recall, and F1 score). The results of 95.8\% accuracy and 95.1\% F1 score illustrate that SAGE$^+$ is proficient in identifying both vulnerable and non-vulnerable samples. GAT utilizes a masked attention mechanism, which helps to concentrate on the specific combination of nodes and edges, resulting in a 95.2\% accuracy and 94.6\% F1 score. Additionally, RGGCN uses gated margins and residuals to handle complex features and vanishing gradient problems, resulting in an accuracy rate of 95.4\% and an F1 score of 94.9\%. UniMP's label information and FiLM's dynamic modulation mechanism are useful for selecting critical features for vulnerability detection, yielding accuracy and F1 score of 95.5\% and 94.8\%, and 95.5\% and 95.0\%, respectively. SAGE is specifically designed to handle massive graphs (IPAGs of large sizes), leading to an accuracy rate of 95.6\% and an F1 score of 94.9\%. Moreover, by taking edge depth into consideration, SAGE$^+$ slightly enhances each metric by approximately 0.2\%.

In terms of stability, GAT has the highest level of stability. The variation in the numerical values of each performance metric (accuracy, precision, recall, and F1 score) across different testing samples is less than 1. In contrast, RGGCN shows a difference of almost 2\% between the minimum and maximum values of each indicator. FiLM, UniMP, SAGE, and SAGE$^+$ fall in an intermediate stability level, with a range of about 1.5\%. Specifically, the accuracy range of SAGE$^+$ is [95.8-0.8, 95.8+0.7], and the F1 score range is [95.1-0.8, 95.1+0.5], which is more stable than SAGE's accuracy and F1 score.

For the Java dataset, the evaluation scores of each model are 2\% higher than the scores on the C dataset. The primary reason for this difference is that the syntax of Java in IPAGs is more detailed than that in C, carrying more structural information. Additionally, the size of the Java dataset is smaller than that of the C dataset. SAGE$^+$ still achieves the best performance among the six models. Moreover, the range of each score for the six models is smaller than that of the C dataset, suggesting that they have stable performance when dealing with Java.

\begin{table*}{}
\scriptsize
\caption{Experimental Results of the C and Java Datasets (\%)}
\begin{center}
\begin{tabular}{|l|C|C|C|C|C|C|C|C|C|C|C|C|C|C|C|C|}
\hline
& \multicolumn{8}{c|}{C Dataset} & \multicolumn{8}{c|}{Java Dataset}\\
\hline
GNNs & \multicolumn{2}{c|}{Accuracy} & \multicolumn{2}{c|}{Precision} & \multicolumn{2}{c|}{Recall} & \multicolumn{2}{c|}{F1 Score} & \multicolumn{2}{c|}{Accuracy} & \multicolumn{2}{c|}{Precision} & \multicolumn{2}{c|}{Recall} & \multicolumn{2}{c|}{F1 Score}\\
\hline
\multirow{2}{*}{GAT} & \multirow{2}{*}{95.2} & +0.3 & \multirow{2}{*}{94.0} & +0.5 & \multirow{2}{*}{95.2} & +0.4 & \multirow{2}{*}{94.6} & +0.4  & \multirow{2}{*}{97.1} & +0.3 & \multirow{2}{*}{97.0} & +0.5 & \multirow{2}{*}{97.2} & +0.4 & \multirow{2}{*}{97.1} & +0.4 \\
&  & -0.4 &  & -0.5 &  & -0.7 &  &  -0.6 &  & -0.4 &  & -0.5 &  & -0.4 &  &  -0.4\\
\hline
\multirow{2}{*}{RGGCN} & \multirow{2}{*}{95.4} & +0.7 & \multirow{2}{*}{94.3} & +0.5 & \multirow{2}{*}{95.4} & +0.5 & \multirow{2}{*}{94.9} & +0.5 & \multirow{2}{*}{97.2} & +0.3 & \multirow{2}{*}{97.3} & +0.3 & \multirow{2}{*}{97.1} & +0.3 & \multirow{2}{*}{97.2} & +0.3  \\
&  & -1.1 &  & -1.1 &  & -1.9 &  &  -1.1 &  & -0.6 &  & -0.6 &  & -0.4 &  &  -0.5 \\
\hline
\multirow{2}{*}{FiLM} & \multirow{2}{*}{95.5} & +0.6 & \multirow{2}{*}{94.9} & +0.3 & \multirow{2}{*}{95.1} & +0.4 & \multirow{2}{*}{95.0} & +0.4 & \multirow{2}{*}{97.2} & +0.4 & \multirow{2}{*}{97.4} & +0.4 & \multirow{2}{*}{97.1} & +0.3 & \multirow{2}{*}{97.3} & +0.4  \\
&  & -0.9 &  & -1.0 &  & -1.0 &  & -1.0 &  & -0.5 &  & -0.6 &  & -0.5 &  &  -0.5 \\
\hline
\multirow{2}{*}{UniMP} & \multirow{2}{*}{95.5} & +0.9 & \multirow{2}{*}{94.8} & +0.5 & \multirow{2}{*}{94.9} & +0.5 & \multirow{2}{*}{94.8} & +0.5 & \multirow{2}{*}{97.3} & +0.4 & \multirow{2}{*}{97.5} & +0.4 & \multirow{2}{*}{97.1} & +0.3 & \multirow{2}{*}{97.2} & +0.4  \\
&  & -1.0 &  & -1.2 &  & -1.1 &  &  -1.1 &  & -0.5 &  & -0.6 &  & -0.5 &  &  -0.5 \\
\hline
\multirow{2}{*}{SAGE} & \multirow{2}{*}{95.6} & +0.8 & \multirow{2}{*}{94.9} & +0.5 & \multirow{2}{*}{95.0} & +0.5 & \multirow{2}{*}{94.9} & +0.5 & \multirow{2}{*}{97.3} & +0.4 & \multirow{2}{*}{97.6} & +0.4 & \multirow{2}{*}{97.2} & +0.3 & \multirow{2}{*}{97.4} & +0.4  \\
&  & -1.0 &  & -1.1 &  & -0.9 &  &  -1.0 &  & -0.5 &  & -0.6 &  & -0.5 &  &  -0.5\\
\hline
\multirow{2}{*}{SAGE$^+$} & \multirow{2}{*}{95.8} & +0.7 & \multirow{2}{*}{95.1} & +0.4 & \multirow{2}{*}{95.2} & +0.6 & \multirow{2}{*}{95.1} & +0.5 & \multirow{2}{*}{97.5} & +0.4 & \multirow{2}{*}{97.7} & +0.4 & \multirow{2}{*}{97.4} & +0.4 & \multirow{2}{*}{97.6} & +0.4  \\
&  & -0.8 &  & -0.9 &  & -0.8 &  &  -0.8 &  & -0.3 &  & -0.4 &  & -0.3 &  &  -0.4\\
\hline
\end{tabular}
\label{table:ResultC}
\end{center}
\end{table*}




\begin{table}
\caption{Results of the Combined Java/C Dataset (\%)}
\begin{center}
\begin{tabular}{|l|c|c|c|c|}
\hline
\textbf{GNN Model} & \textbf{Accuracy} & \textbf{Precision} & \textbf{Recall} & \textbf{F1}\\

\hline
GAT & 95.6 & 94.7 & 95.3 & 94.9\\
\hline
RGGCN & 95.7 & 94.9 & 95.5 & 95.2 \\
\hline
FiLM & 95.7 & 95.5 & 95.1 & 95.3  \\
\hline
UniMP & 95.8 & 95.3 & 95.7 & 95.6\\
\hline
SAGE & 96.1 & 95.4 & 95.7 & 95.6 \\
\hline
SAGE$^+$ & 96.3 & 95.5 & 95.8 & 95.7\\
\hline
\end{tabular}
\label{table:ResultJavaAndC}
\end{center}
\end{table}

\subsubsection{Performance on Individual CWEs}

\begin{figure*}[h]
    \centering
    \includegraphics[width=1\textwidth, height =4cm] {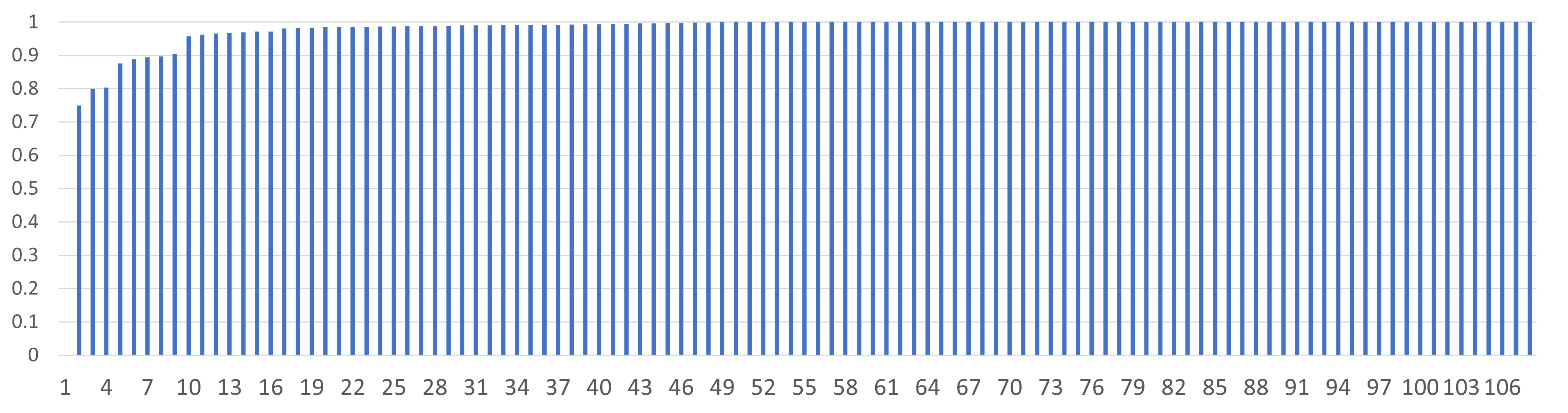}
    \caption{Accuracy for Individual Vulnerabilities}
    \label{fig:ResultOfEachCWE}
\end{figure*}

We employed the high-performance SAGE$^+$-based model for assessing the performance of each CWE. The results are presented in Fig. \ref{fig:ResultOfEachCWE}. Notably, the accuracy for CWE-562 (index 1) is observed to be 0, indicating that the model struggles to classify this specific vulnerability. CWE-562 involves a function returning the address of a stack variable, leading to unintended program behavior, often resulting in a crash. In the C programming language, the '\&' operator is utilized to retrieve the memory address where a variable is stored. In our methodology, we treat the combination of the '\&' operator and the variable name as a single token, encoding it through a pre-trained model. This approach poses challenges for the model in distinguishing instances where a function returns the address of a stack variable. Consequently, the model encounters difficulty in accurately classifying such cases. 

The accuracy for CWE-480, CWE-364, and CWE-451 (index 2-4) is below 80\%. CWE-480 is attributed to the use of an incorrect operator. The vulnerable feature associated with this CWE is often confined to a single node, making detection challenging. CWE-364 arises from race conditions, wherein program behavior hinges on the relative timing of events, such as the order in which threads or processes execute. The inherent complexity of race conditions presents challenges in detection. CWE-451 involves user interface misrepresentation of critical information. Defining misrepresentation solely based on semantic meaning is intricate, adding complexity to the detection process. The accuracy for CWEs (index 5-9) is close to 90\%, which is considered acceptable. The accuracy for the remaining 99 CWEs surpasses the model's overall performance.

\subsection{Effectiveness of Graph Reduction}

\begin{table}
\caption{Compression Ratios}
\begin{center}
\begin{tabular}{|c|c|c|c|}
\hline
\textbf{Efficiency} & \textbf{IPAGs without} & \textbf{IPAGs} & \textbf{Reduction}\\
\textbf{Metrics} & \textbf{Reduction} & \textbf{} & \textbf{Ratio}\\
\hline
Number of Nodes & 37,274,276 & 22,511,155 & 39.6\%\\
\hline
Number of Edges & 58,138,339 & 43,380,491 & 25.4\%\\
\hline
Time (min/10 Epochs) & 51 & 30 & 41.2\% \\
\hline
Storage (GB) & 281.3 & 171.1 & 48.1\%\\
\hline
\end{tabular}
\label{table:CompressionRatios}
\end{center}
\end{table}

Table \ref{table:CompressionRatios} shows the compression ratios on four metrics after compressing sequence and aggregation structures in IPAGs. Before reduction, totally, there were 37 million nodes and 58 million edges. After compressing sequence and aggregation structures, with a 39.6\% reduction in the number of nodes and a 25.4\% decrease in edges. In addition, the time for training sees a 41.2\% decrease per 10 epochs, indicating remarkable speed improvements. Additionally, storage requirements shrink by 48.1\%, signifying significant space savings. 

\begin{figure}[ht]
\begin{tabular}{ccccc}
\includegraphics[scale=0.35]{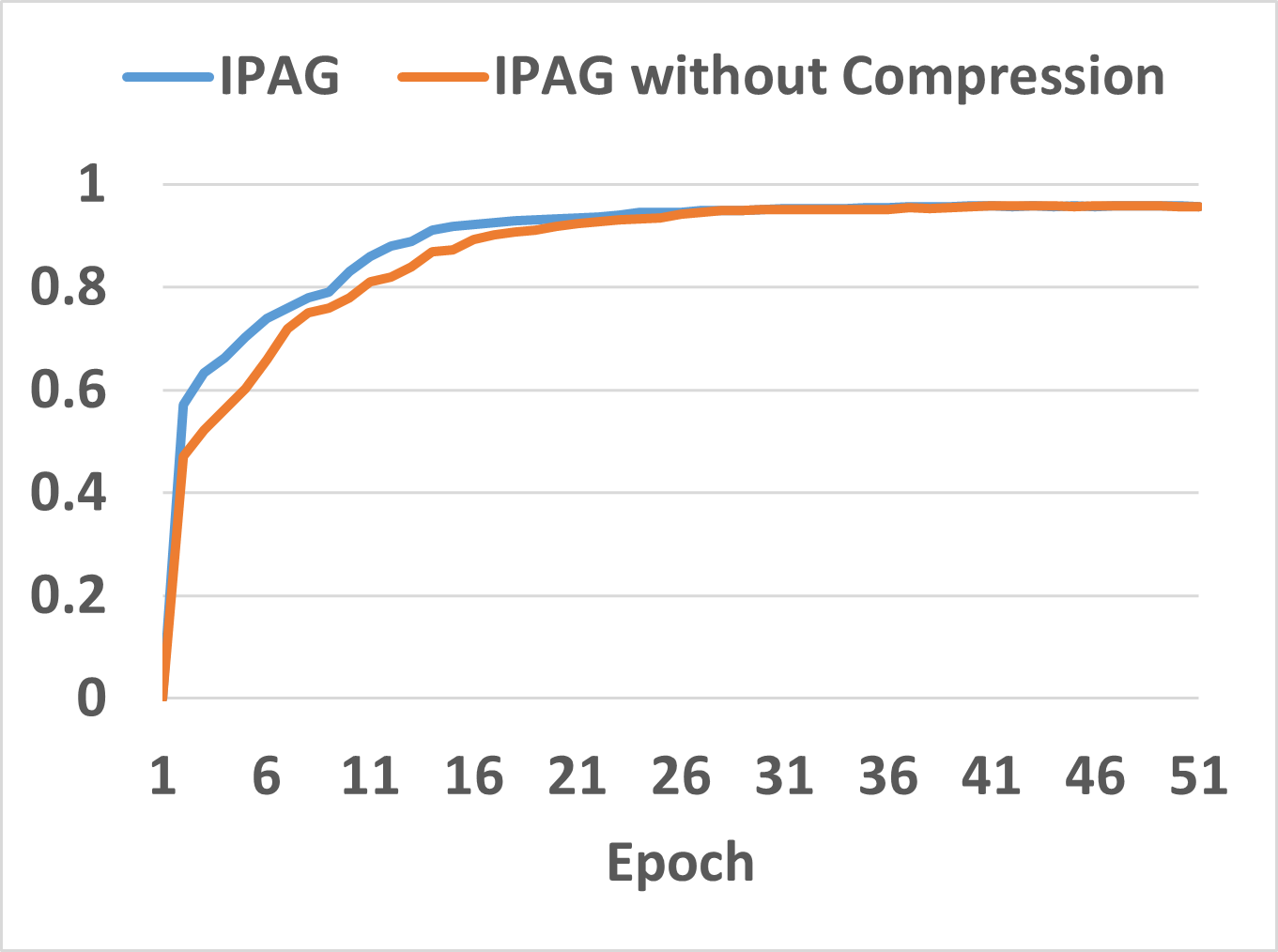}
&
\includegraphics[scale=0.35]{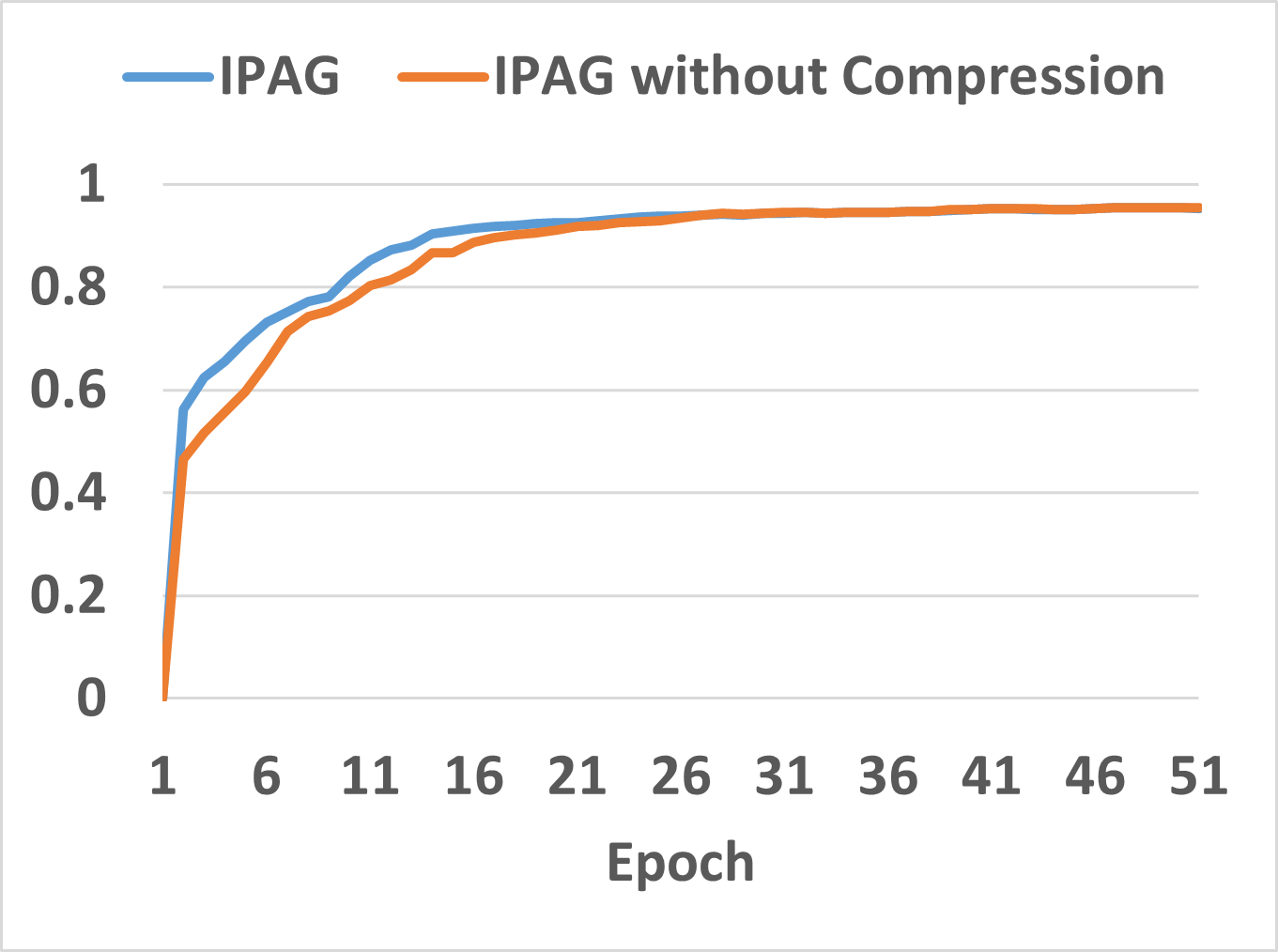} \\
(a) Accuracy  & (b) Precision\\
\includegraphics[scale=0.35]{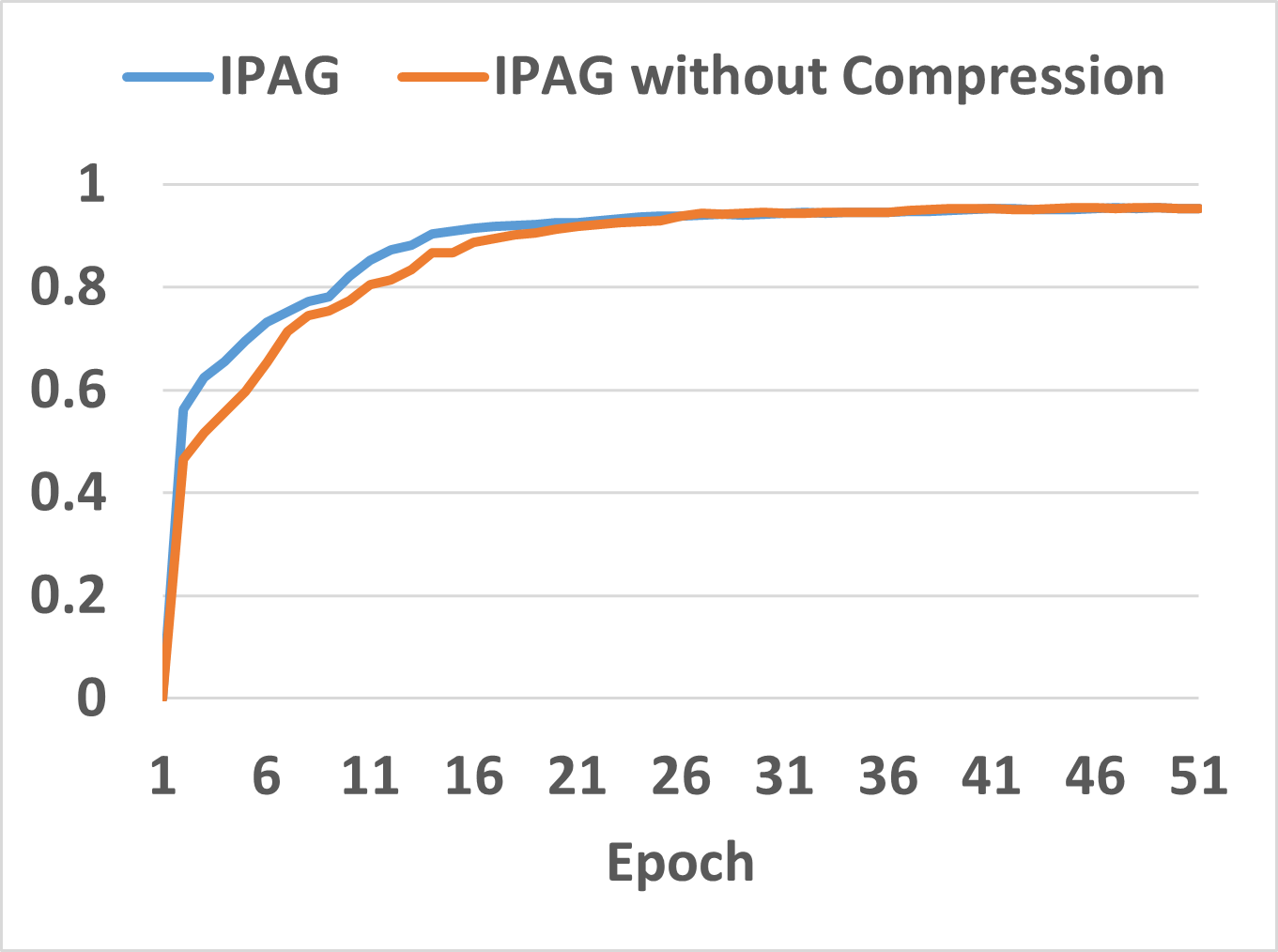} 
&
\includegraphics[scale=0.35]{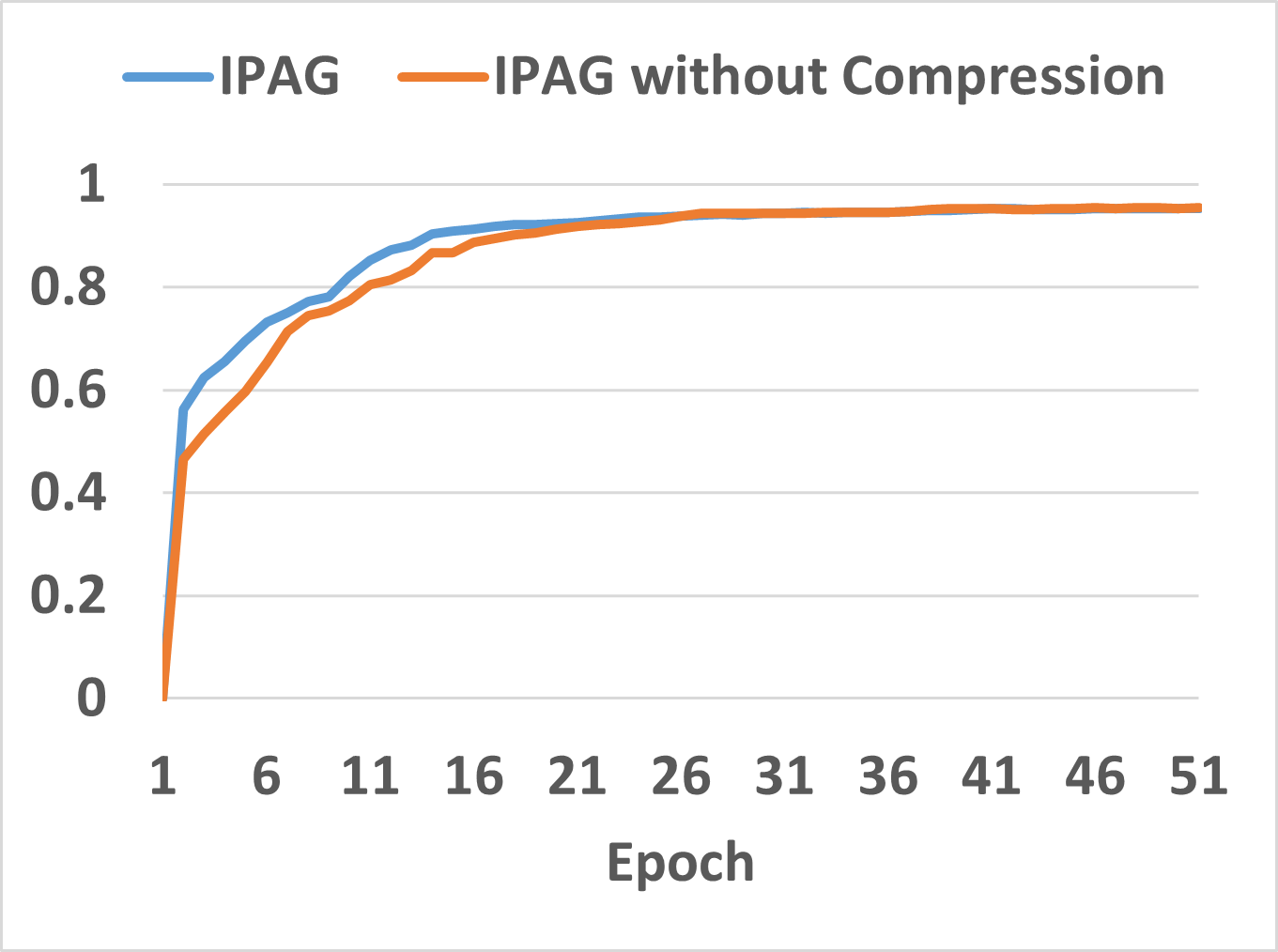}\\
(c) Recall  & (d) F1 Score\\
\end{tabular}
\caption{Comparison of IPAGs without Reduction and IPAGs}
\label{fig:ReductionComparison}
\end{figure}

We also apply IPAGs and IPAGs without compression on SAGE$^+$ with heterogeneous training, and the comparison results are shown in Fig. \ref{fig:ReductionComparison}. After epoch 30, both models achieve optimum, and they have similar accuracy, precision, recall, and F1 scores. The IPAGs also make learning faster for the same learning rate.

\subsection{Comparison with Other Graph Representations}

We have compared IPAGs to other commonly used source code representations for program analysis, including ASTs, CFGs, and PDGs. ASTs are tree-based representations of the abstract syntactic structure of source code. CFGs represent the source code of a program as a graph, with nodes representing basic blocks of code and edges representing the flow of control between them. Each basic block is a sequence of non-branching instructions that end with a branching instruction, such as a conditional jump or a subroutine call. PDGs are a graph-based representation of the dependencies between statements in the source code of a program. Each node in the graph corresponds to a statement, and edges represent the dependencies between them, such as data dependencies, control dependencies, or other types of dependencies.

\subsubsection{Overall Performance}

\begin{table}
\caption{Comparison of Different Graph Representations (Accuracy\%)}
\begin{center}
\begin{tabular}{|c|c|c|c|c|c|c|}
\hline
\textbf{Graph} & \textbf{GAT} & \textbf{RGGCN} & \textbf{FiLM} & \textbf{UniMP} & \textbf{SAGE} & \textbf{SAGE$^+$}\\

\hline
AST & 84.3 & 84.4 & 84.6 & 84.6 & 84.7 & 84.9\\
\hline
PDG & 86.7 & 86.8 & 86.8 & 86.9 & 87.2 & 87.3\\
\hline
CFG & 90.1 & 90.2 & 90.1 & 90.4 & 90.7 & 90.7 \\
\hline
IPAG & \textbf{95.2} & \textbf{95.4} & \textbf{95.5} & \textbf{95.5} & \textbf{95.6} & \textbf{95.8}\\
\hline
\end{tabular}
\label{table:GRComparison}
\end{center}
\end{table}

Table \ref{table:GRComparison} shows the results of the comparison. IPAGs have demonstrated significantly better performance than ASTs, CFGs, and PDGs with an average accuracy of 95\%. On the other hand, AST-based models perform poorly in vulnerability detection as they solely capture the features of abstract syntactic structure. The six GNN models have an average accuracy of 84\%. PDG-based models achieved an average accuracy of 87\% by incorporating features of dependencies between statements, while CFG-based models utilized information on code blocks and flow to achieve an average accuracy of 90\%.


\subsubsection{Performance on Samples with Call Relations}

\begin{table}
\caption{Comparison of Different Graph Representations on Call Samples (Accuracy\%)}
\begin{center}
\begin{tabular}{|c|c|c|c|c|c|c|}
\hline
\textbf{Graph} & \textbf{GAT} & \textbf{RGGCN} & \textbf{FiLM} & \textbf{UniMP} & \textbf{SAGE} & \textbf{SAGE$^+$}\\

\hline
AST & 73.1 & 73.2 & 73.1 & 73.6 & 73.8 & 73.9\\
\hline
PDG & 75.2 & 75.4 & 75.3 & 75.7 & 75.9 & 75.9\\
\hline
CFG & 88.2 & 88.2 & 88.7 & 88.5 & 88.9 & 88.9 \\
\hline
IPAG & \textbf{97.1} & \textbf{97.2} & \textbf{97.5} & \textbf{97.4} & \textbf{97.6} & \textbf{97.8}\\
\hline
\end{tabular}
\label{table:GRComparisonCALL}
\end{center}
\end{table}

The comparison results shown in Table \ref{table:GRComparisonCALL} only take into account the positive and negative samples with call relations. The findings indicate that IPAGs outperform ASTs, CFGs, and PDGs. Notably, all six IPAG-based models achieved an accuracy of over 97\%. However, the performance of the other three graphs (ASTs, CFGs, and PDGs) on call relations is lower than their overall performance.

\subsubsection{Ablation Study on IPAGs}

\begin{figure}[h]
    \centering
    \includegraphics[width=0.45\textwidth, height=4cm] {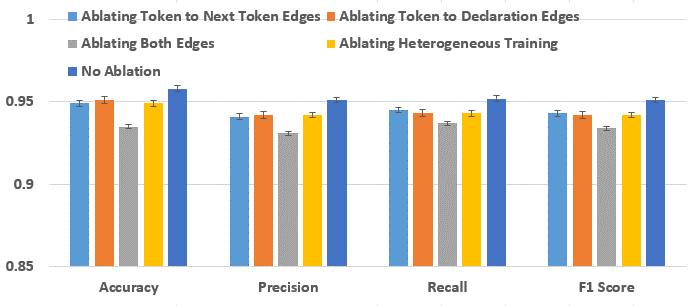}\\
    \caption{The Results of Ablation Study on C Dataset}
    \label{fig:Ablation}
\end{figure}

Fig. \ref{fig:Ablation} shows the experimental results of the ablation study on IPAGs. Ablating token-to-next-token edges leads to an approximate 1\% decrease in all measured metrics, with the most substantial impact on the F1 Score. Ablating token-to-declaration-edges also causes a decrease in all metrics, but the impact is slightly less severe compared to the removal of token-to-next-token edges. When both types of edges are removed, there's a about 2\% reduction in all metrics. In terms of the training method, substituting heterogeneous training with homogeneous training causes about a 1\% decrease across all metrics. The model performs best when no ablation is applied, indicating that both edge types and heterogeneous training play a crucial role in the model's performance.


\subsection{Comparison with Other Related Works}

This section compares our approach with several other methods, including VulDeePecker \cite{Li2018}, $\mu$VulDeePecker \cite{zou2019}, Luo et al. \cite{luo2022Journal}, DEVIGN \cite{zhou2019devign}, Lin et al. \cite{lin2018cross}, FUNDED \cite{Wang2020}, DeepWuKong\cite{cheng2021deepwukong}. Our previous experiments demonstrated that, among the six GNN models in the HAGNN framework, SAGE$^+$ delivered the best performance. Therefore, we denote SAGE$^+$ within the HAGNN framework as HAGNN-SAGE$^+$ and use it as the representative model for our comparative study.

VulDeePecker utilizes a BiLSTM neural network to detect vulnerabilities by extracting code gadgets based on library/function calls. $\mu$VulDeePecker enhances VulDeePecker by incorporating the dependency feature to detect more types of vulnerabilities. Luo et al. treat source code as text and use a fine-tuned BERT model to detect integer overflow errors. DEVIGN uses GNN models to learn features through an AST variant, while FUNDED uses a GNN to operate on a graph representation with multiple types of edges based on ASTs and combines them through a GRU. DeepWuKong utilizes GATs, GCNs, and k-GNNs on sliced PDGs with labels assigned at the statement level. We reproduce this approach by applying PDGs with labels at the routine level specifically on GCNs.

\begin{figure*}[h]
    \centering
    \includegraphics[width=1\textwidth, height =4cm] {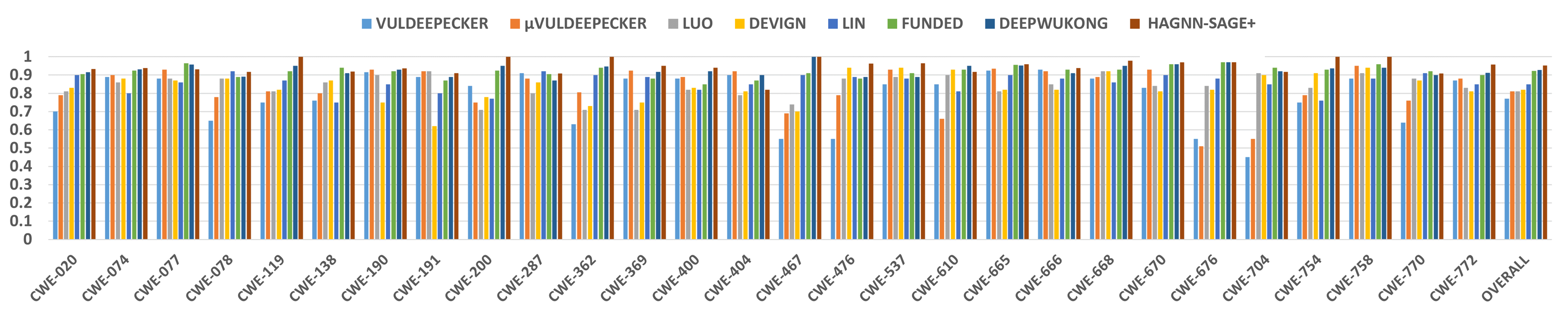}
    \caption{Comparison of Individual Vulnerabilities in the TIFS Dataset}
    \label{fig:ResultOfRelatedWork}
\end{figure*}

Our comparative study uses the TIFS dataset from FUNDED, which includes 38,845 negative and 34,035 positive samples and covers 28 types of vulnerabilities in C programs. Fig. \ref{fig:ResultOfRelatedWork} displays the comparison results. Overall, HAGNN-SAGE$^+$ exhibits the best performance. DeepWuKong and FUNDED, being the latest work, achieved nearly 93\% accuracy, a 3\% improvement over previous works but 2\% less than IPAG-based models. The accuracy of the remaining works ranged between 80\% and 85\%.

Fig \ref{fig:ResultOfRelatedWork} shows various methods for detecting different types of vulnerabilities. The BiLSTM-based models, VULDEEPECKER and uVULDEEPECKER, have high accuracy for some vulnerabilities, like CWE-190, CWE-191, and CWE-665, but their performance is poor for other types. Luo's method works well only for CWE-190, but its accuracy drops to 70\% when applied to other vulnerabilities. Lin's method achieves good accuracy for some vulnerabilities, but HAGNN-SAGE$^+$ achieves better or equal scores for most. DEVIGN has high accuracy only for a few vulnerabilities. FUNDED and DeepWuKong have similar performance, demonstrating effectiveness across various vulnerability types. However, they both face challenges in addressing vulnerabilities classified under CWE-191, CWE-400, and CWE-404. Our approach outperforms all methods, achieving over 90\% accuracy for most vulnerability types, except for CWE-404, which has an accuracy of 82\%. This vulnerability often occurs in a sequence and requires related feature capturing, which graph-based methods are not good at, whereas VULDEEPECKER and VULDEEPECKER with BiLSTM perform better.

\section{Application}

We have applied the resultant models of the heterogeneous training to four real-world open-source projects \cite{luo2022}. They are listed in Table \ref{table:Application}. The project sizes range from 112 KLOC (thousand lines of code) to 615 KLOC. In total, these projects have 1,464 KLOC, 12,438 classes, and 74,351 routines, with 2,186 of these routines being vulnerable.

\begin{table}
\caption{The Applications}
\begin{center}
\begin{tabular}{|l|r|r|r|r|}
\hline
\textbf{Project}  &  \textbf{KLOC} & \textbf{Classes}  & \textbf{Routines} & \textbf{\#Vul.} \\
\hline
FFmpeg  & 615 & 3,478 & 36,505 & 637 \\
\hline
OpenSSL & 361 & 2,203 & 19,922 & 619 \\
\hline
WireShark & 112 & 863 & 7,123 & 555 \\
\hline
GNU-preg & 201 & 1,112 & 10,801 & 375 \\
\hline
Total  & 1,464 & 12,438 & 74,351 & 2,186\\
\hline
\end{tabular}
\label{table:Application}
\end{center}
\end{table}

Table \ref{table:ApplicationResult} presents the results of evaluating six HAGNN models. All six models achieved high scores on accuracy (ranging from 93.3\% to 95.2\%) and F1-scores (ranging from 91.8\% to 93.8\%). A high accuracy score indicates that the models performed well overall on all routines, and a high F1 score suggests good performance on both vulnerable and non-vulnerable routines. SAGE$^+$ model had the highest scores for all four metrics, while GAT had the lowest FPR and SAGE$^+$ had the lowest FNR. All models had FPRs lower than 1\%, indicating a low chance of misclassifying a non-vulnerable routine as vulnerable. Compared to the training results in Tables \ref{table:ResultC} and \ref{table:ResultJavaAndC}, the performances on real-world applications dropped but remain promising. In particular, the SAGE$^+$ model is outstanding.

Additionally, we have applied our approach to 966 real-world vulnerable C programs, covering 487 Linux Kernel CVEs since 2018 and 479 security patches of over 200 open-source GutHub projects with high star ratings and commits. To obtain the vulnerable source code of the Linux Kernel CVEs, we manually pinpoint the associated vulnerable routines from older Linux Kernel versions according to the CVE reports. These vulnerabilities primarily stem from memory management issues, such as buffer overflows, use-after-free errors, null pointers, race conditions in multithreaded code, and improper input validation. To obtain the vulnerable source code from the patched GitHub projects, we leverage a pre-trained expert model \cite{liu2021combining} to analyze the code commits for vulnerability patches. When a patch is identified, we use the code changes to trace the previous version of the patched routine. 

Using HAGNN-SAGE$^+$, we successfully detected 825 of the 996 vulnerabilities, achieving an accuracy of 85.4\%. One group of undetected vulnerabilities involves device drivers and third-party module for the kernel's extensive use in diverse environments. The other group are related to use-after-free errors and race conditions, which require precise modeling of memory lifetimes, interactions with other routines  and non-deterministic threads.

\begin{table}
\caption{Prediction Results for Real-World Applications (\%)}
\begin{center}
\begin{tabular}{|l|c|c|c|c|c|c|}
\hline
\textbf{GNN Model} & \textbf{A} & \textbf{P} & \textbf{R} & \textbf{F1} & \textbf{FPR} & \textbf{FNR}\\

\hline
RGGCNs & 93.3 & 88.7 & 95.2 & 91.8 & 0.5 & 22.3\\
\hline
FiLM & 93.6 & 88.6 & 95.5 & 91.9 & 0.4 & 22.6\\
\hline
UniMP & 93.8 & 89.2 & 95.9 & 92.4 & 0.5 & 21.2 \\
\hline
GAT & 94.7 & 90.9 & 95.1 & 93.0 & 0.9 & 17.0\\
\hline
SAGE & 94.9 & 90.6 & 96.2 & 93.4 & 0.3 & 18.2\\
\hline
SAGE$^+$ & 95.2 & 91.1 & 96.7 & 93.8 & 0.2 & 17.7 \\
\hline
\end{tabular}
\label{table:ApplicationResult}
\end{center}
\end{table}

\section{Related Work}

We review related work on machine learning-based detection of code vulnerability in terms of different representations of code: (a) code as graphs, (b) code as trees, and (c) code as text.

Graph-based methods typically capture the structural characteristics of various types of graphs, such as control-flow graphs (CFGs), data-flow graphs (DFGs), and data dependence graphs (DDGs), using techniques such as graph neural networks (GNNs), convolutional neural networks (CNNs), or recurrent neural networks (RNNs), to construct models for detecting vulnerabilities. VulDeePecker \cite{Li2018} models C/C++ source code as a graph, leveraging graph-level and node-level features for vulnerability prediction. $\mu$VulDeePecker \cite{zou2019} is an enhanced version incorporating data and control dependencies from system dependency graphs (SDGs). Including code attention and localized information enables the model to effectively detect 40 different types of vulnerabilities. DeepTective \cite{rabheru2022hybrid}  combines gated recurrent units (GRUs) and graph convolutional networks (GCNs) to leverage both syntactic and semantic information for detecting SQLi, XSS, and OSCI vulnerabilities in PHP code. Velvet \cite{ding2022velvet} integrates graph-based and sequence-based neural networks to capture the local and global context of code and identify vulnerable patterns. DEVIGN \cite{zhou2019devign} uses a joint graph representation of code snippets by merging ASTs, CFGs, and DFGs for detecting vulnerabilities in C programs with high accuracy. FUNDED \cite{Wang2020} creates a method-level graph representation using nine types of edges to ASTs, and uses a pre-trained word2vec network updated by a GRU for node embeddings, achieving an average accuracy of 92\%. HGVul \cite{song2022hgvul} utilizes the code property graph combined with natural code sequence as a graph representation, which is fed into three different GNN models, GCN, GAT, and GGNN, achieving an F1 score of 88.3\%. EL-VDetect \cite{sun2023enhanced} integrates serialization-based and graph-based neural networks with an attention mechanism to capture code semantics and achieves 90.72\% accuracy on a real-world dataset. Wang et al. \cite{liu2021combining} utilize contract graphs for smart contract vulnerability detection, achieving an average accuracy of 89\%. AMPLE \cite{wen2023vulnerability} is a vulnerability detection framework with graph simplification and enhanced graph representation learning that aims to capture global vulnerable information. Cao et al. \cite{cao2022mvd} propose a statement-level vulnerability detection approach based on flow-sensitive graph neural networks (FS-GNN) to capture implicit memory-related vulnerability patterns. DeepWuKong \cite{cheng2021deepwukong} is another statement-level vulnerability detector that utilizes GATs, GCNs, and k-GNNs on sliced PDGs. These methods utilize GNN models to detect vulnerabilities, but their capacity to model diverse relationships between code elements contributing to different vulnerabilities is limited.

Tree-based approaches extract features by traversing the ASTs and their variants. Dong et al. \cite{dong2018defect} utilized code sequences extracted from ASTs as semantic features and the frequency as token features to build a fully connected neural network for detecting vulnerabilities in Android binary executables. Wang et al. \cite{wang2016} preserved three types of nodes, and converted them into code sequences. These sequences are mapped into high-dimension vectors to train deep belief networks (DBNs) for detecting software weakness. POSTER \cite{lin2017} and Lin et al. \cite{lin2018cross} discover vulnerabilities in the function level by building a bidirectional LSTM network with ASTs-based sequences. Dam et al. \cite{dam2017automatic} built a sequence to sequence the LSTM network to learn the semantic and syntactic features from ASTs in Java methods. SySeVR \cite{li2021sysevr} divided programs into small pieces and generated multiple representations from ASTs to exhibit the syntax and semantics characteristics of vulnerabilities. These approaches are incapable of capturing intricate program structural properties (branches or parallel statements).

Text-based methods exploit NLP models for vulnerability detection by treating source code as text. Peng et al. \cite{peng2015building} used n-gram models and Wilcoxon rank-sum optimization on Java source code vectors. Hovsepyan et al. \cite{hovsepyan2012} and Pang et al. \cite{pang2015} used SVM with bag-of-words (BOW) and n-grams representations of Java code. Scandariato et al. \cite{scandariato2014} used text-mining for vulnerability prediction. Lee et al. \cite{lee2017learning} developed an Instruction2vec model on assembly codes. Yamaguchi et al. \cite{yamaguchi2011} applied NLP approaches to API symbol prediction. Luo et al. \cite{luo2022Journal} used natural language syntax with BERT for Java code. Russell et al. \cite{Russell2018} used CNNs and RNNs on embedded source representations. Le et al. \cite{le2018maximal} proposed a sequential autoencoder for feature extraction on binary code. Sestili et al. \cite{sestili2018towards} used one-hot vectors and memory networks for buffer overflow detection. These methods focus on semantic information but may overlook the structural features. Shabtai et al. \cite{shabtai2009} applied principal component analysis to the ASTs of source code to identify vulnerable code. Similarly, Mokhov et al. \cite{mokhov2015} used numerous methods from WEKA \cite{holmes1994weka} and the ASTs as characteristics to construct prediction models.

In addition to vulnerability detection, there are other source code representations for program analysis. CodeBERT \cite{feng2020codebert} is a bimodal pre-trained model by adding programming language to base BERT \cite{Devlin2018}, which supports downstream NL-PL applications. SourcererCC \cite{sajnani2016sourcerercc} transformed programs into regularized token sequences and ordered by an optimized inverted index for code clone detection. Based on program CFGs and PDGs, Allamanis et al. \cite{allamanis2017learning} applied Gated Graph Neural Networks to predict variable names and detect variable misuses, and DeepSim \cite{zhao2018deepsim} encoded flows into a semantic matrix for measuring code functional similarity.

\section{Conclusion}

We have presented IPAGs as a novel source code representation for predicting software vulnerabilities with heterogeneous attention GNN models. IPAGs consist of three node types and six edge types that capture a variety of source code characteristics. The heterogeneous training employs six message-passing units that update node features based on related edge types, while a global attention mechanism identifies significant features linked to vulnerabilities. Using two large datasets containing 220 types of vulnerabilities in Java and C, the study demonstrates that the approach can achieve high accuracy. The resulting GNN models also show promising results in identifying more than 2,100 vulnerabilities from real-world software projects. Furthermore, the comparative study reveals that our approach outperforms existing machine-learning methods for detecting vulnerabilities.  

While the experiments conducted in this study primarily concentrated on C and Java programs, it is important to note that the proposed approach is applicable to other programming languages. This versatility arises from the fact that IPAGs serve as a language-agnostic representation of source code. In our future work, we plan to extend this study to encompass the detection of code vulnerabilities in other commonly used programming languages, such as C\#, C++, and Python.

Furthermore, to provide more valuable insights for identifying and addressing bugs, we aim to investigate multi-class classification to categorize specific CWEs. Additionally, we aspire to apply GNN models to identify previously unknown vulnerabilities in real-world applications.

\section*{Acknowledgments}
This work was supported in part by US National Science Foundation (NSF) under grants 1820685 and 2101118.

\bibliographystyle{IEEEtran}  
\bibliography{References}  

\vspace{11pt}


\vfill

\end{document}